\documentclass[letterpaper]{sig-alternate}

\newfont{\mycrnotice}{ptmr8t at 7pt}
\newfont{\myconfname}{ptmri8t at 7pt}
\let\crnotice\mycrnotice
\let\confname\myconfname

\permission{Permission to make digital or hard copies of all or part of this work for personal or classroom use is granted without fee provided that copies are not made or distributed for profit or commercial advantage and that copies bear this notice and the full citation on the first page. Copyrights for components of this work owned by others than ACM must be honored. Abstracting with credit is permitted. To copy otherwise, or republish, to post on servers or to redistribute to lists, requires prior specific permission and/or a fee. Request permissions from permissions@acm.org.}
\conferenceinfo{MobiHoc'15,}{June 22--25, 2015, Hangzhou, China.}
\copyrightetc{Copyright \copyright~2015 ACM \the\acmcopyr}
\crdata{978-1-4503-3489-1/15/06\ ...\$15.00.\\
DOI: http://dx.doi.org/10.1145/2746285.2755969
}

\usepackage{url}
\usepackage{flushend}  
\usepackage{graphicx} 
\usepackage{times}    
\usepackage{url}      
\usepackage{balance}
\usepackage{color}
\usepackage{wrapfig}
\usepackage{caption}
\usepackage{subcaption}
\usepackage{array}
\usepackage{marginnote}
\usepackage{epstopdf}
\usepackage{xcolor}
\usepackage{amsfonts}
\usepackage{amsmath,amssymb}
\usepackage{epstopdf}
\usepackage{cite}
\usepackage{hyperref}

\DeclareMathOperator*{\argmax}{arg\,max}

\def \sys {\textit{UbiBreathe}}

\begin{document}

\title{UbiBreathe: A Ubiquitous non-Invasive \\WiFi-based Breathing Estimator}

\numberofauthors{3} 
\author{
\alignauthor
Heba Abdelnasser\\
       \affaddr{Computer and Sys. Eng. Dep.}\\
       \affaddr{Alexandria University}\\
        \email{\large{heba.abdelnasser@alexu.edu.eg}}
\alignauthor
Khaled A. Harras\\
       \affaddr{Computer Science Dep.}\\
       \affaddr{Carnegie Mellon University}\\
       \email{\large{kharras@cs.cmu.edu}}
\alignauthor
Moustafa Youssef\\
       \affaddr{Wireless Research Center}\\
       \affaddr{E-JUST}\\
       \email{\large{moustafa.youssef@ejust.edu.eg}}
}

\maketitle

\begin{abstract}
Monitoring breathing rates and patterns helps in the diagnosis and potential avoidance of various health problems. Current solutions for respiratory monitoring, however, are usually invasive and/or limited to medical facilities. In this paper, we propose a novel respiratory monitoring system, \sys{}, based on ubiquitous off-the-shelf WiFi-enabled devices. Our experiments show that the received signal strength (RSS) at a WiFi-enabled device held on a person's chest is affected by the breathing process. This effect extends to scenarios when the person is situated on the line-of-sight (LOS) between the access point and the device, even without holding it. \sys{} leverages these changes in the WiFi RSS patterns to enable ubiquitous non-invasive respiratory rate estimation, as well as apnea detection.

We propose the full architecture and design for \sys{}, incorporating various modules that help reliably extract the hidden breathing signal from a noisy WiFi RSS.  The system handles various challenges such as noise elimination, interfering humans, sudden user movements, as well as detecting abnormal breathing situations. Our implementation of \sys{} using off-the-shelf devices in a wide range of environmental conditions shows that it can estimate different breathing rates with less than 1 breaths per minute (bpm) error. In addition, \sys{} can detect apnea with more than 96\% accuracy in both the device-on-chest and hands-free scenarios. This highlights its suitability for a new class of anywhere respiratory monitoring.

\end{abstract}


\section{Introduction}
Monitoring breathing rates is an important predictor of a number of serious problems such as cardiac arrests, strokes, or chronic obstructive pulmonary diseases \cite{parkes2011rate,de2009influence}. Existing breathing monitors used in hospitals typically require special devices attached to the human body (e.g. a mask or a nasal cannula). These special devices are usually annoying for patients, limit their movement, and more importantly are not appropriate for remote patient monitoring. The camera on a mobile phone has also been used to measure the respiration rate of a user by analyzing the user's chest motion \cite{poh2010non}. This approach, however, requires a certain amount of light to work properly, and therefore, cannot for instance be used to monitor a sleeping infant's breathing rate in a dark room. In addition, in other scenarios, the mobile phone camera usage would quickly consume the battery power, and raise privacy concerns. Recently, a number of RF-based techniques, e.g. \cite{droitcour2009signal,ballal2012breathing,chen2008human, lazaro2010analysis,patwari2014breathfinding} 
have been proposed for contact-free breathing monitoring. Nevertheless, such systems have a limited range, high cost, and/or high deployment overhead.

\begin{figure}[!t]
\centering
        \begin{subfigure}[t]{0.48\textwidth}
                \centering
                \includegraphics[width=\textwidth]{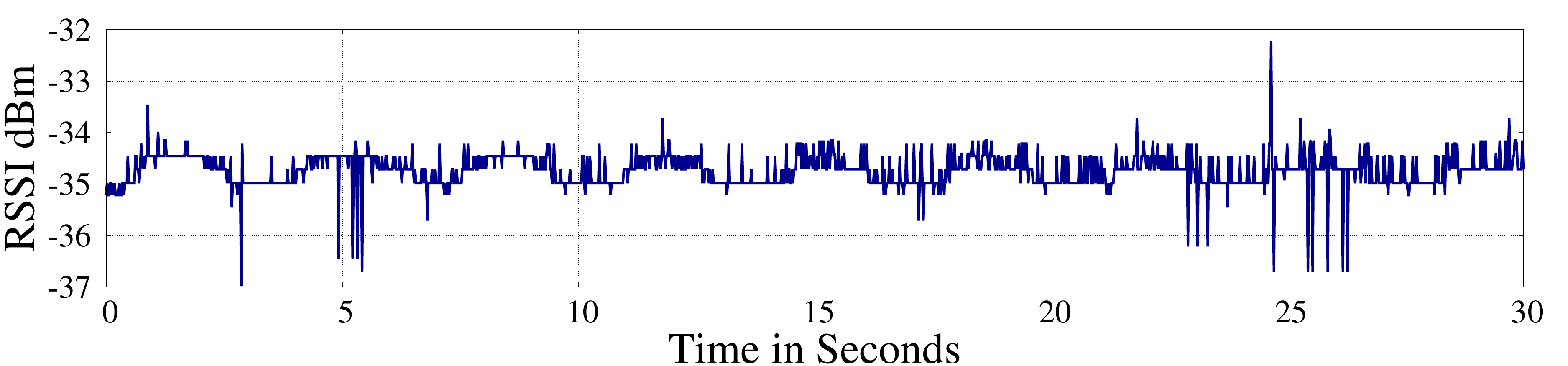}
                \caption{Raw measured WiFi signal strength.}
                \label{fig:raw_}
        \end{subfigure}

	\begin{subfigure}[t]{0.48\textwidth}
                \centering
                \includegraphics[width=\textwidth]{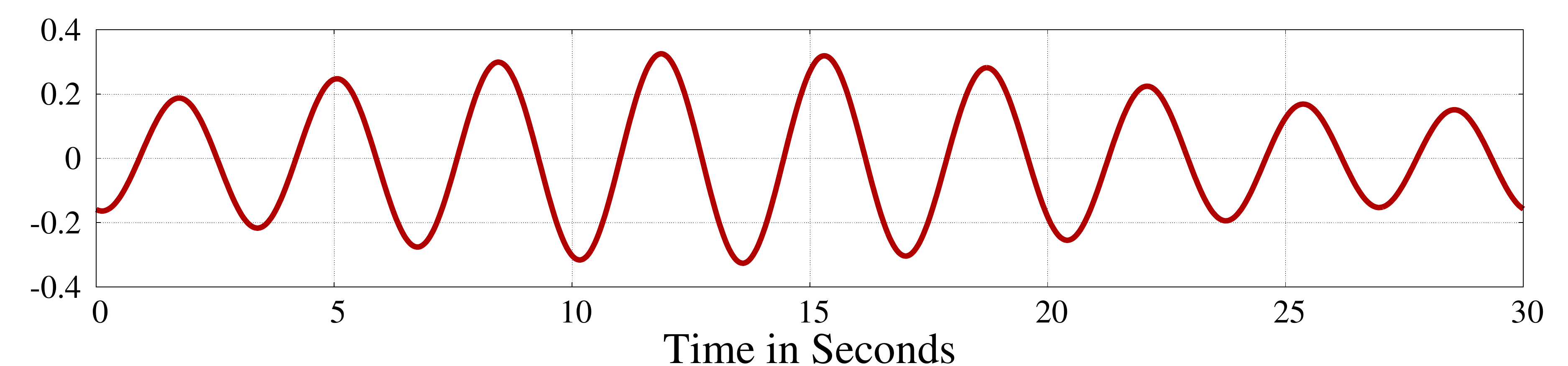}
                \caption{The breathing waveform after processing by \sys{}.}
                \label{fig:feq_mag_filtering}
        \end{subfigure}

\caption{\sys{} can extract the embedded breathing signal from the noisy signal strength. The ground truth breathing rate is 18 bpm.}
\label{fig:magnitude_thresholding_effect}
\vspace{-0.2in}
\end{figure}

In this paper, we present \sys{}: A system that provides affordable, pervasive, non-intrusive and easy to use/deploy respiration monitoring \footnote{\textbf{A video showing the \sys{} system in action can be watched at: }\texttt{\href{https://www.youtube.com/watch?v=MnlKz2n5mrI&list=PL5hKa_dp7Ry83sbtfrjz5WZFj1ekSTgXZ}{UbiBreathe youtube video.}}}. 
\sys{} is a software-only solution that can work with any WiFi-enabled device without the need of any special hardware, can monitor multiple persons in parallel, detect breathing anomalies, and display the full breathing signal in realtime. The basic idea \sys{} leverages is that the chest/lungs are large organs, and the inhaling and exhaling motion of a breathing person causes a dominant periodic component in the received WiFi signal at a receiver positioned on the user's chest. This ``modulated'' WiFi signal due to the breathing process can be analyzed to extract different useful information about the person's breathing pattern. Moreover, this effect extends to the scenario when the person is on the line-of-sight (LOS) between the AP and WiFi-enabled device, without carrying any device. This is particularly useful in applications such as in-home sleep apnea diagnostic systems for the elderly, or monitoring a baby's respiration while sleeping to avoid Sudden Infant Death Syndrome (SIDS), which is the leading cause of death among infants one month to one year old \cite{mckenna2005babies}.

We present the \sys{} system architecture along with the details of how it can extract the breathing signal and different respiratory information from the noisy WiFi received signal strength (RSS), while handling different practical challenges. An example of a typical RSS input and resulting output from our \sys{} system is shown in Figure \ref{fig:magnitude_thresholding_effect}. To obtain such output, \sys{} employs a number of processing modules to reduce the effect of interfering humans, handle sudden user movement, and detect outliers. Moreover, \sys{} analyzes the extracted breathing signal in realtime and raises alarms about the loss of the breathing signal, indicating the presence of apnea.

We implement \sys{} using standard WiFi APs, laptops, and cell phones, and evaluate its performance in two different environments including a typical apartment, and one floor in our engineering building, under varying environmental parameters. Overall, our results show that \sys{} can achieve a high breathing rate detection accuracy of less than 1 breaths per minute (bpm) error for different breathing rates. In addition, it can detect apnea with more than 92\% accuracy and less than 10\% false positive and false negative rates based on the readings of a single AP. This accuracy further increases to 96\% and less than 4\% false positive and false negative rates based on the RSS of five overheard APs.

In summary, our contributions are three-fold:
\begin{itemize}
\item We present the design and architecture of a non-intrusive system for estimating breathing rates and detecting apnea using off-the-shelf WiFi-enabled devices. The system works in both device-on-chest and hands-free scenarios.
\item We present the details of different signal processing modules to handle a number of practical scenarios covering noise reduction, human interference, sudden user movement, and signal outliers.
\item We implement the system on different Android devices and thoroughly evaluate it in two different testbeds under different environmental conditions, interfering users, and operational scenarios.
\end{itemize}

The remainder of this paper is organized as follows. Section~\ref{related_work} provides an overview on related work. Section~\ref{system} presents the architecture and overview of \sys{} followed by a detailed description of various system components in Section~\ref{componets}. The evaluation setup and system assessment in real environments are detailed in Section~\ref{evaluation}. Finally, we conclude and discuss future work in Section~\ref{conclusion}.

\section{Related Work}
\label{related_work}

We break down prior work related to our system into recent breathing monitoring systems and the state-of-the-art WiFi-based event detection systems.

\subsection{Respiration Monitoring Systems}

Respiration monitoring systems often used in hospitals typically require special \textit{contact-based} devices attached to the human body. Capnometers \cite{mogue1988capnometers}, for example, are widely used in practice and require patients to have a mask or nasal cannula constantly attached to them. Photoplethysmography (PPG) \cite{shariati2005comparison}, an optical technique used to detect blood volume changes in the microvascular tissues by illuminating the skin and measuring light absorbtion changes, is also used by hospitals in ICUs via sensors attached to a patient's finger (e.g. pulse oximeters) to monitor breathing and heart rates. These solutions, however, use special hardware, may be annoying to the user, limit her movement, and may not be suitable for remote patient monitoring at home.

Along the same line, recent studies have demonstrated the feasibility of measuring heart rates by placing the index finger over a cellphone camera with its flash turned on or through special heart rate sensors (e.g. as in the Samsung Galaxy S5 phone). The camera or sensor records the light absorbed by the finger tissue, and from that video, each frame is processed by splitting each pixel into RGB components whose values can be used to acquire a PPG signal \cite{scully2012physiological, jonathan2010investigating, pelegris2010novel}. Mobile phones have also been shown to measure the respiration rate of a user by analyzing the chest motion \cite{poh2010non}. These camera-based or sensor-based solutions share the drawbacks of consuming a lot of the scarce phone energy, requiring a certain amount of light to work properly (therefore not suitable for monitoring a person sleeping in a dark room), requiring special sensors, requiring direct contact with the user skin, and/or raising privacy concerns.

Due to skin sensitivity and other issues that inhibit attaching sensors to the body, \textit{contact-free} RF respiration monitoring devices were proposed including microwave doppler radars 
\cite{droitcour2009signal,ballal2012breathing}, ultra-wideband (UWB) radars 
\cite{chen2008human,lazaro2010analysis}, and ISM-based systems \cite{patwari2014breathfinding,kaltiokallio2014non}. These systems can provide high accuracy in respiration rate detection due to the special frequency band used and/or dense device deployment. However, their  main drawbacks are the limited range of high frequency devices used, as well as their high cost. In addition, \cite{kaltiokallio2014non} uses special ZigBee devices with high gain directional antennas.

\sys{}, on the other hand, provides a \textbf{\emph{software-only solution}} that uses off-the-shelf WiFi devices installed on a ubiquitous scale and available in various user devices. It can provide accurate estimation with only one AP-device pair. However, it has to deal with the challenges associated with the WiFi signal characteristics and low device density.

\subsection{WiFi-based Event Detection Systems}
With the ubiquity of WiFi-enabled devices and infrastructure, various WiFi-based event detection systems have been proposed. For example, device-free activity recognition relying on RSSI fluctuations, or the more detailed channel state information (CSI) in standard WiFi networks, have emerged as ubiquitous solutions for presence detection \cite{el2010propagation,kosba2012robust,kosba2012rasid, youssef2007challenges, saeed2014ichnaea, abdel2013monophy,Kafrawy:PIMRC11,seifeldin2010deterministic,eleryan2011aroma, sabek2012multi,aly2013new}
, tracking individuals \cite{kosba2009analysis, seifeldin2013nuzzer,Sabek:TRSPOT12,sabek2014ace}
, recognition of different human activities \cite{ding2011rftraffic,sigg2013rfa,al2012rf,Kassem:VTC12}, and gesture recognition systems \cite{pu2013whole,abdelnasser2015wigest}.

\sys{} extends these systems to a new domain: ubiquitous health-care. It monitors breathing signal information of multiple users simultaneously (each user needs to carry her own device), and provides apnea detection capabilities.

\section{System Overview}
\label{system}

The goal of \sys{} is to monitor the breathing of a person by exploiting resulting fluctuations in standard WiFi signals using off-the-shelf mobile devices. The basic idea \sys{} leverages is that a dominant periodic component is introduced in the received WiFi signal as a result of the volume change in the chest/lungs when a person is inhaling and exhaling. Therefore, the WiFi RSS can be analyzed to extract different useful information about the person's breathing pattern. For simplicity of illustration, we present the current and following sections assuming that the mobile device is held on the person's chest. However, the same principles apply to the hands-free mode of operation, where the person is away from the mobile device but in the line-of-sight between a transmitter and a receiver (e.g. an AP or any other WiFi-enabled device). We quantify the system accuracy for both configurations in Section~\ref{evaluation}.

\begin{figure}[!t]
\centering
\includegraphics[width=3.3in]{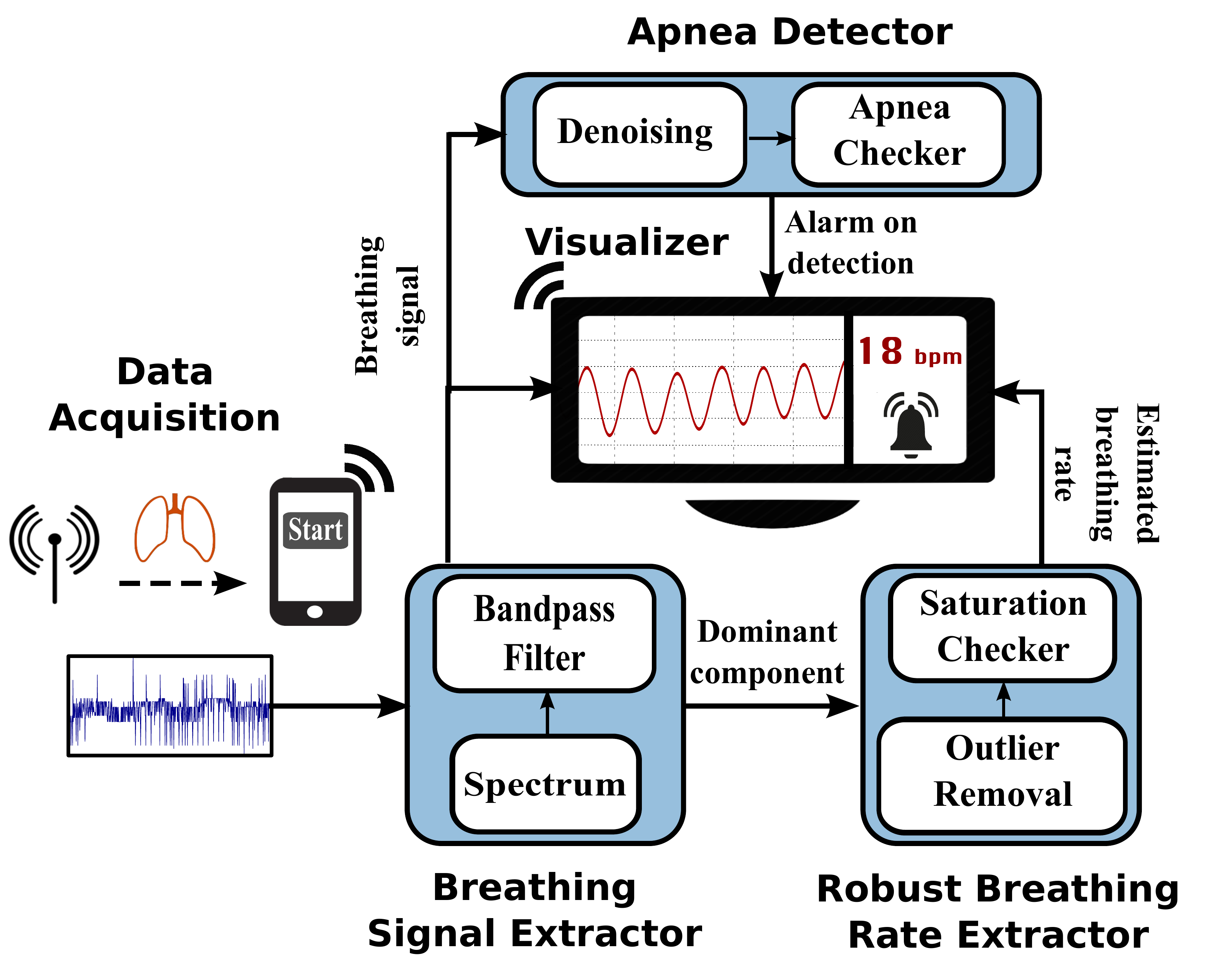}
\caption{\sys{} architecture. \sys{} has four components: Breathing Signal Extractor, Robust Breathing Rate Extractor, Apnea Detector, and Realtime Visualizer modules.}
\label{fig:arch}
\vspace{-0.1in}
\end{figure}

The architecture of our system, depicted in Figure \ref{fig:arch}, includes four major modules: the Breathing Signal Extractor, the Robust Breathing Rate Extractor, the Apnea Detector, and the Realtime Visualizer. All modules can be run on the user device, offloaded to the cloud, or to a local user device (e.g. a nearby laptop) \cite{mtibaa2013towards,mtibaa2013towardsmcc}. The process starts when the user activates the breathing estimation process through the system GUI on her phone. This trigger-based activation helps in reducing the system's energy consumption, as it is not running all the time; this activation also partially reduces the effect of interference. The data acquisition module on the user's mobile device then begins collecting the WiFi received signal strength (RSS) in realtime.

Overall, the system modules operate as follows. The RSS is first processed by the \emph{Breathing Signal Extractor} module that filters the noise from the input signal and extracts the breathing signal from sliding windows over the input RSS stream as well as the instantaneous breathing rate in realtime. The extracted breathing signal is then passed to the \emph{Robust Breathing Rate Extractor} module that filters outliers and provides a more stable signal reading (e.g. for the patient's chart) by fusing the different estimates over successive windows. The \emph{Apnea Detector} module applies further denoising techniques to the breathing signal and accordingly checks for the absence of the breathing pattern. Finally, the \emph{Realtime Visualizer} module combines the output of the different modules in a user friendly visual output and raises alarms when an apnea is detected.

We note that different information can be dispatched to different user devices. For example, the full information (full breathing signal, instantaneous rate, and apnea alarm) can be displayed on the user's nearby laptop or sent over the Internet for remote monitoring by a health-care facility or a close family member. Alternatively, a single time reading can be displayed on the user's cell phone after a stable reading is obtained (indicated by an audible beep); or an apnea alarm can be sent to the cell phone of the parent of a monitored sleeping child.

\newpage
\section{The UbiBreathe System}
\label{componets}

In this section, we discuss the details of the four modules of the \sys{} system shown in Figure \ref{fig:arch}.

\subsection{Breathing Signal Extractor}
The goal of this module is to extract the full breathing signal from the received noisy WiFi RSS values. We first describe the base operational case with a single steady user and then discuss how \sys{} handles interfering humans in the environment and sudden actions from these users. The output from this module is fed to all other modules simultaneously for different purposes as described in Section~\ref{system}.   

\subsubsection{Basic Operation}
Figure~\ref{fig:feq_filtering} shows the Fast Fourier Transform (FFT) for the WiFi signal in Figure~\ref{fig:magnitude_thresholding_effect}. The person's breathing rate is $18$ breaths per minute (bpm) (i.e. 0.3 Hz). The figure shows that the frequency spectrum of the WiFi signal influenced by the breathing person has a strong component close to the actual breathing rate. This behavior is consistent over different experiments, as we show later in the evaluation section. 

\sys{} leverages this observation to estimate the breathing rate. In particular, we begin by obtaining the frequency spectrum by applying the FFT to a sliding window of the WiFi raw signal of length $n$ samples taken over $W$ seconds\footnote{The effect of these parameters on system performance is analyzed in Section~\ref{evaluation}.}. A band pass filter is then applied to limit the frequencies to those within the range of the normal human breathing rates \cite{lindh2013delmar} (red frequencies in Figure~\ref{fig:feq_filtering}). The band pass cut-off frequencies are therefore set from $0.1$ to $0.5$Hz, which correspond to breathing rates between $6$ and $30$bpm. The breathing rate ($\hat{r}$) is finally estimated as the frequency with the maximum magnitude in the human breathing rate range. More formally:

\begin{figure}[!t]
\centering
      \includegraphics[width=0.5\textwidth]{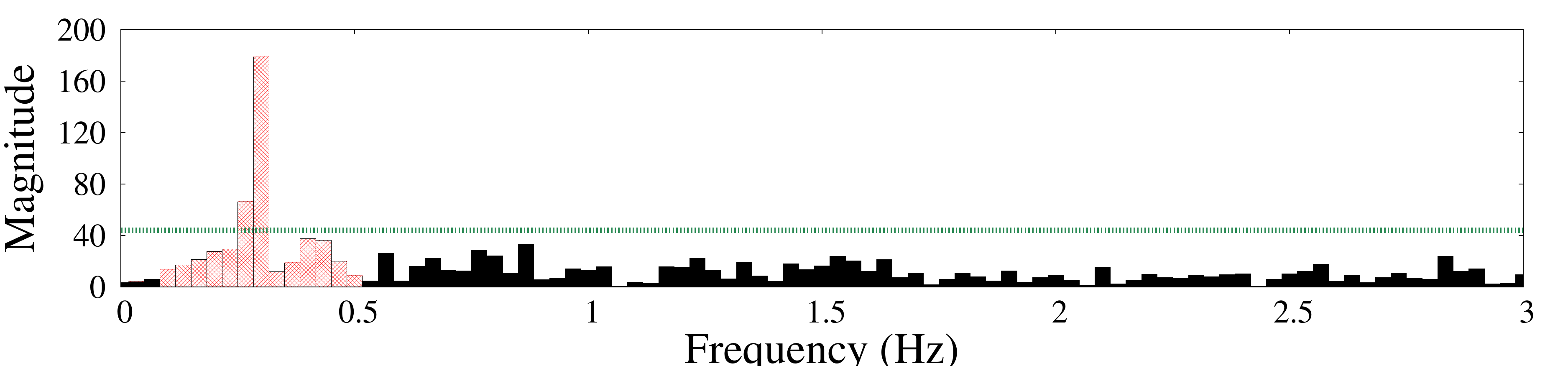}
      \caption{The FFT of the raw WiFi signal in Figure~\ref{fig:magnitude_thresholding_effect}. A band pass filter (red frequencies) is applied to the signal spectrum to limit it to the human breathing range (6 to 30 bpm) and remove the DC component. Another filtering stage based on magnitude/energy thresholds is further applied to the band-limited spectrum (horizontal line).}
                \label{fig:feq_filtering}
\vspace{-0.1in}            
\end{figure}

The breathing rate ($\hat{r}$) is finally estimated as the frequency with the maximum magnitude in the human breathing rate range. More formally:
\begin{equation}
\hat{r} = \argmax_{r_{min} \leq r \leq r_{max}} |\textrm{FFT}(x_{1...n})|
\end{equation}

where $r_{min}$ is the minimum human respiration rate, $r_{max}$ is the maximum human respiration rate, and $x_{1...n}$ are the RSS values in the current sliding window.

This \textbf{\emph{instantaneous}} estimated rate is passed to the next system modules for further processing. To reconstruct the breathing signal, noting that the dominant frequency components in a signal have high energy, we first apply another noise reduction step by trimming all frequencies with low energy, i.e. whose amplitude falls below 25\% of the amplitude of the dominant breathing frequency (green line in Figure~\ref{fig:feq_filtering}). Finally, an inverse FFT operation is performed to obtain the time domain breathing signal (Figure~\ref{fig:feq_mag_filtering}), which is passed to the visualization module.

\subsubsection{Handling Sudden Changes}

\begin{figure}[!t]
\centering
        \begin{subfigure}[t]{1.645in}
            \centering
            \begin{minipage}{\linewidth}
                \includegraphics[width=1\linewidth, height = 0.3\textheight, keepaspectratio=true]{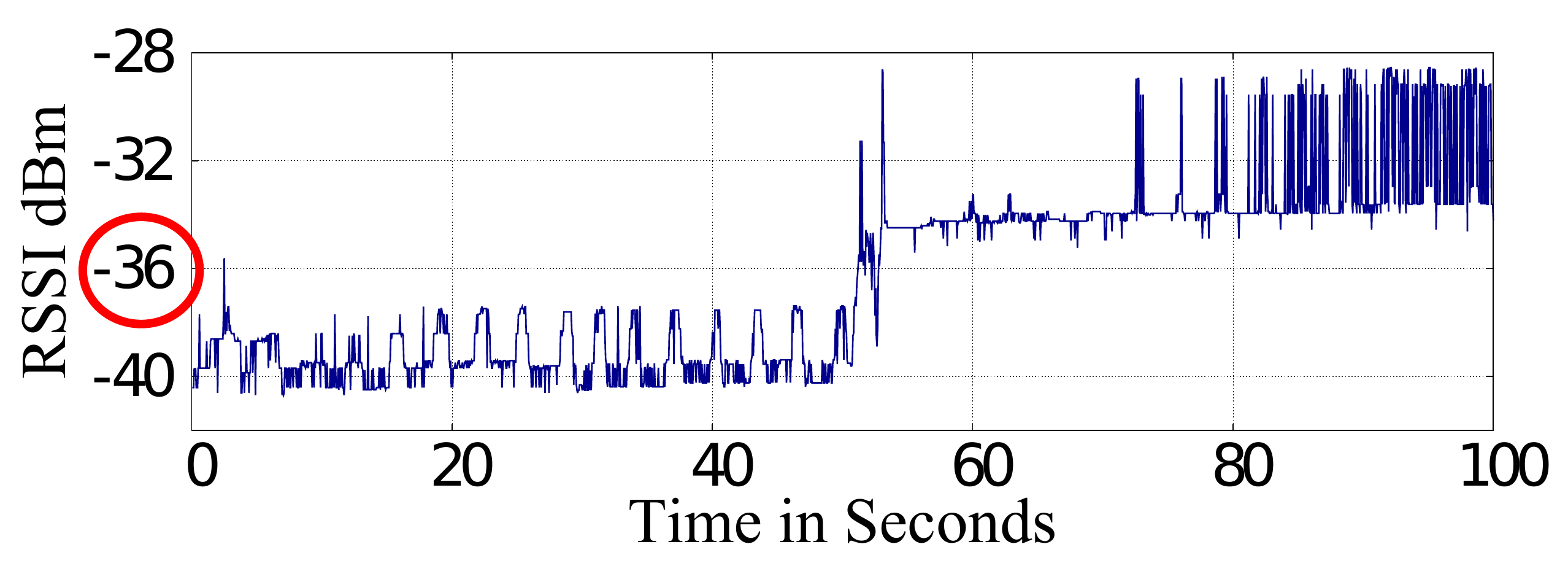}
                \caption*{}

                \includegraphics[width=1\linewidth, height = 0.3\textheight, keepaspectratio=true]{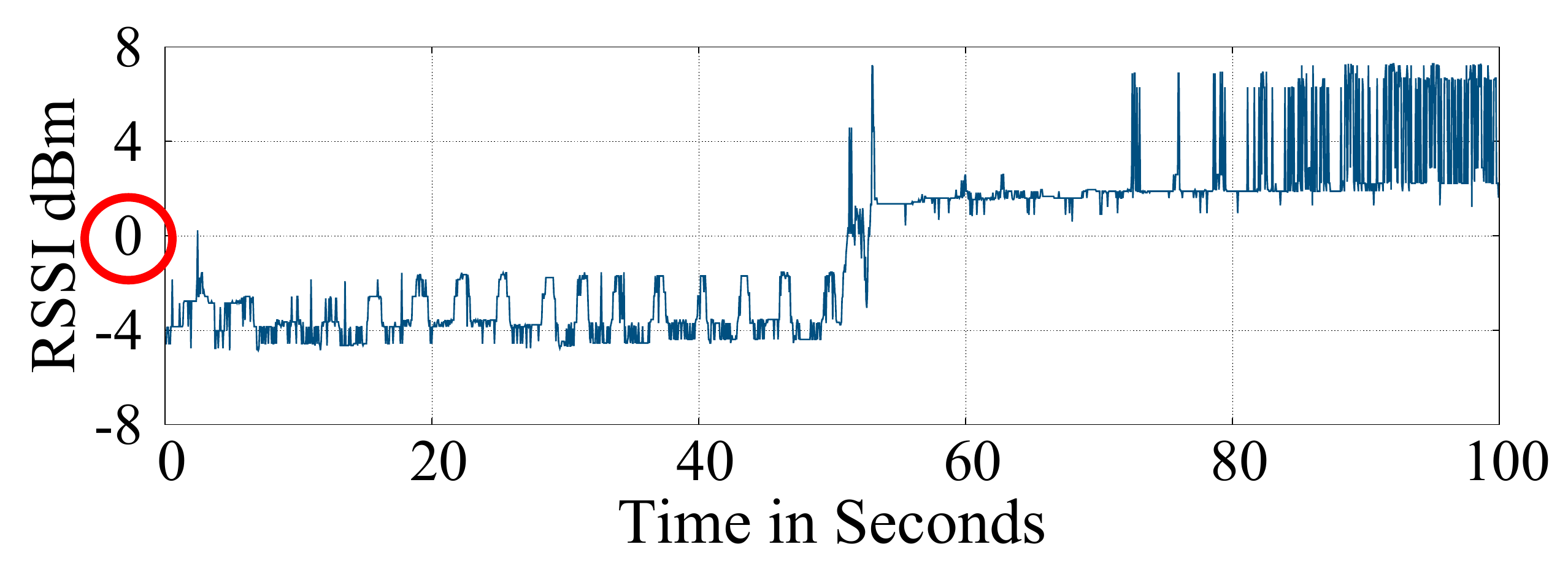}
                \caption*{Signal after full-window mean (DC component) removal. }

                \includegraphics[width=1\linewidth, height = 0.3\textheight, keepaspectratio=true]{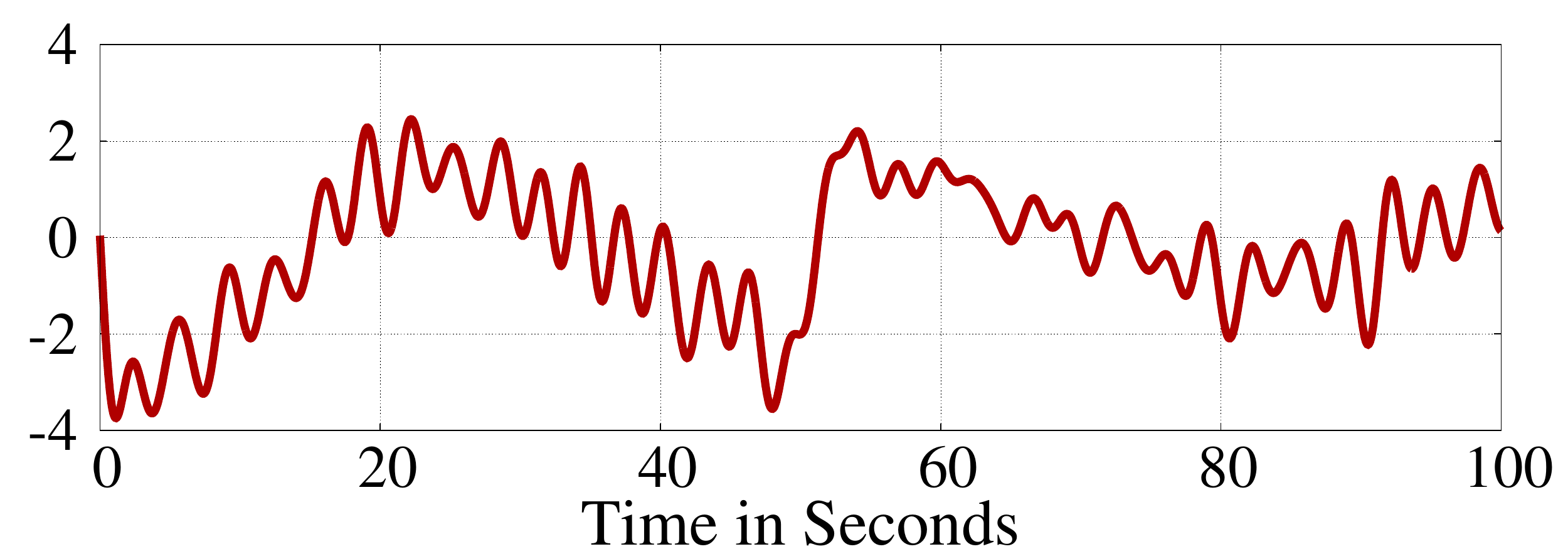}
                \caption*{}

                \includegraphics[width=1\linewidth, height = 0.3\textheight, keepaspectratio=true]{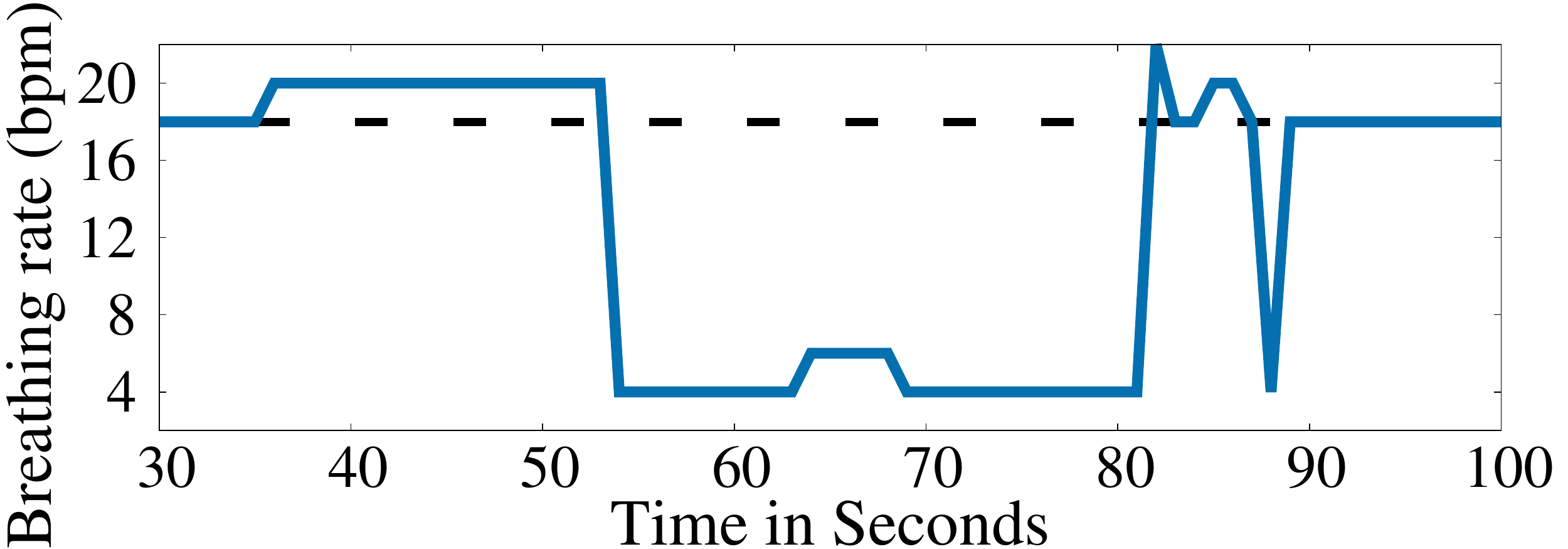}
                \caption*{}

            \end{minipage}
             \caption{Simple mean removal.}
             \label{fig:simple_mean_removal}
        \end{subfigure}
        \begin{subfigure}[t]{1.645in}
            \centering
            \begin{minipage}{\linewidth}
                \includegraphics[width=1\linewidth, height = 0.3\textheight, keepaspectratio=true]{figures/motion_interf_mean_removal/raw_breathing_motion_interf}
                \caption*{\hspace{-4.5cm}Raw signal (duplicated).}

                \includegraphics[width=1\linewidth, height = 0.3\textheight, keepaspectratio=true]{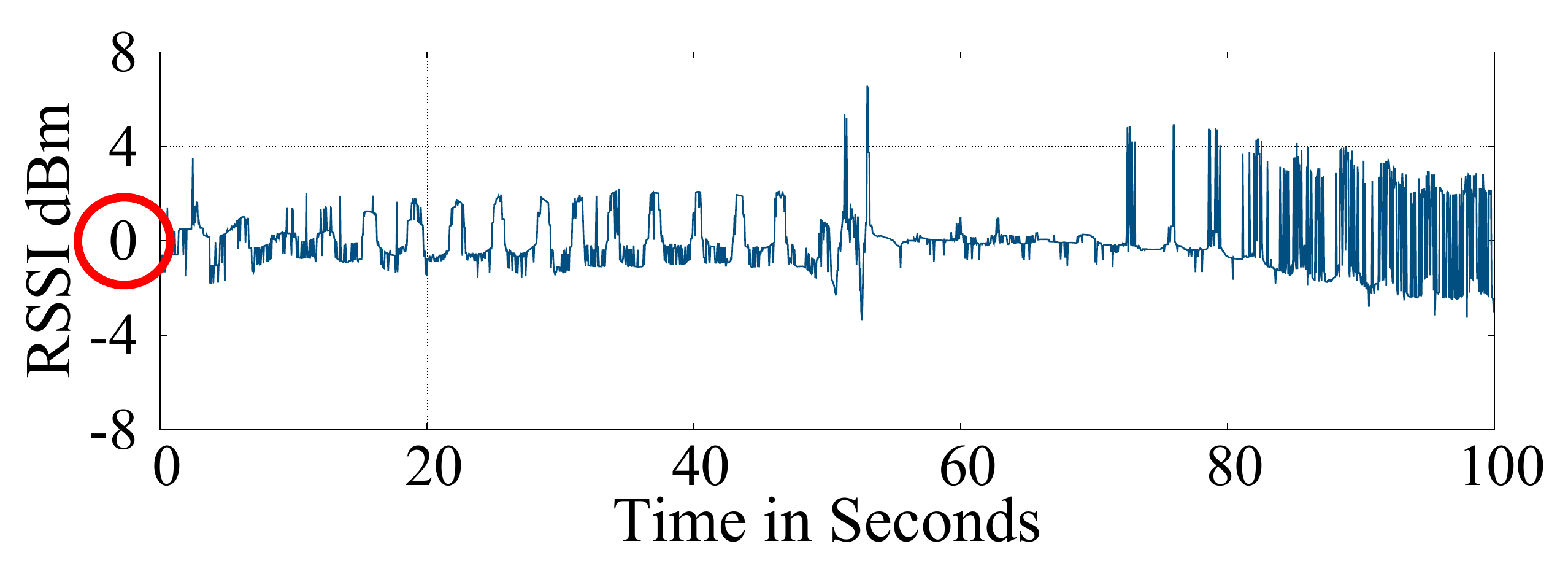}
	       \caption*{Signal after short-window local mean removal.}

                \includegraphics[width=1\linewidth, height = 0.3\textheight, keepaspectratio=true]{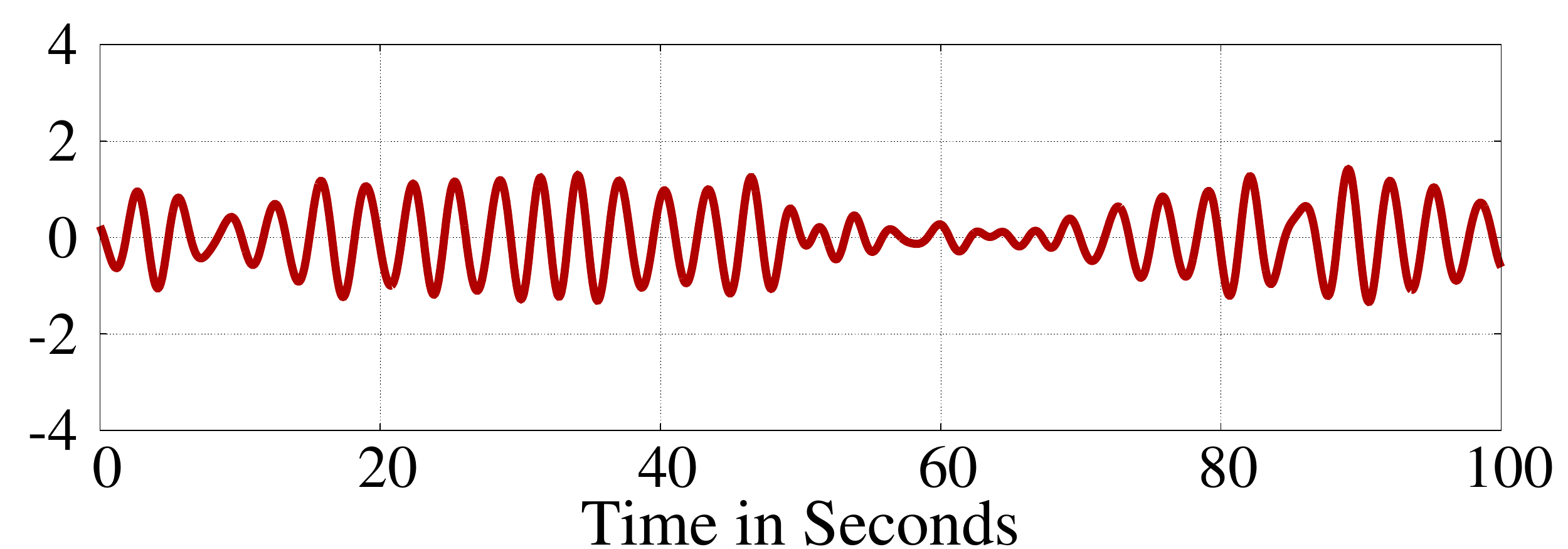}
	       \caption*{\hspace{-4.5cm}Corresponding breathing signal.}
	
                \includegraphics[width=1\linewidth, height = 0.3\textheight, keepaspectratio=true]{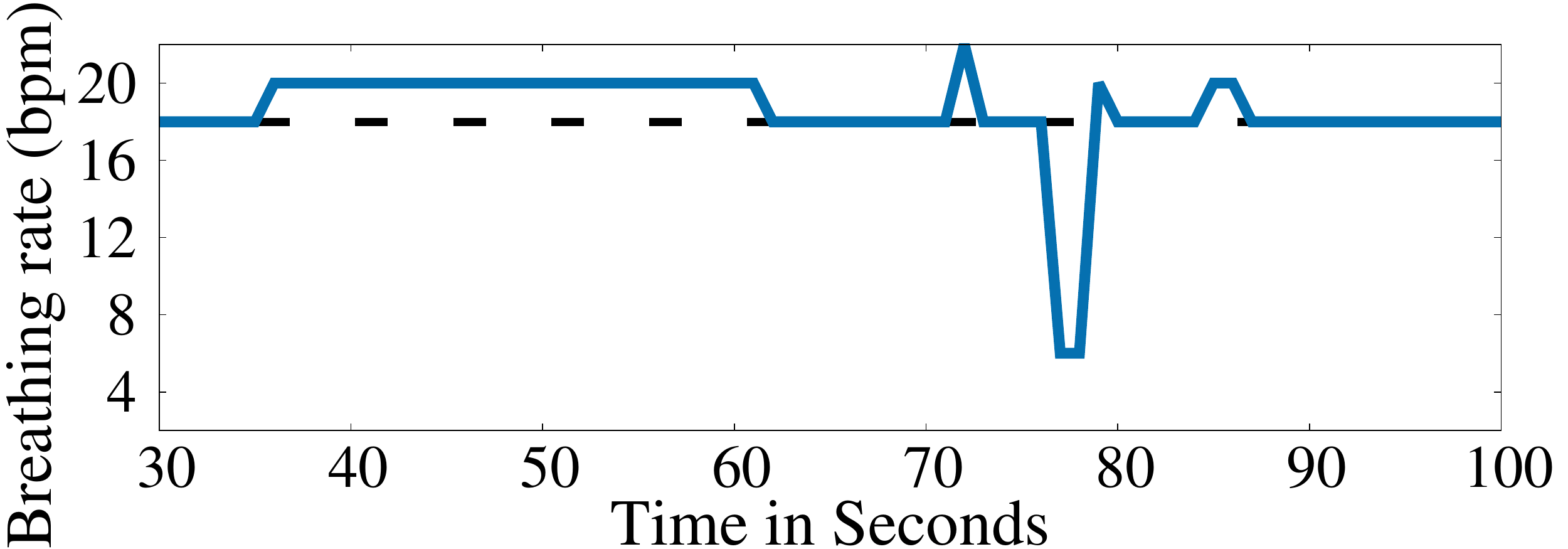}
                \caption*{\hspace{-4.25cm}\mbox{Corresponding estimated instantaneous breathing rate.}}
            \end{minipage}
            \caption{\textbf{Proposed} sub-window local mean removal.}
            \label{fig:sub_mean_removal}
            \end{subfigure}
\caption{Different mean removal techniques for handling sudden changes from the user or other interfering humans. Ground truth breathing rate is shown as a horizontal dotted black line. A simple full-window mean (i.e. DC component) removal does not remove the effect of the sudden change at time $t=52$ sec which affects the estimated breathing rate for an extended period. The proposed sub-window local mean removal technique balances the raw input signal and reduces the sudden change effect.}
\label{fig:dc_removal}
\vspace{-0.1in}
\end{figure}

Sometimes the user may make a sudden action/move or other users may interfere within the line of sight. These actions can lead to large sudden changes in the RSS, and hence the breathing signal as shown in Figure~\ref{fig:dc_removal}. A simple mean removal (i.e. DC component removal) does not solve this issue as shown in Figure~\ref{fig:simple_mean_removal}, making the impact of such actions span a long duration as it appears in multiple overlapping FFT windows. To reduce this effect, we apply a \textbf{\emph{within-window local}} mean removal technique. Specifically, we subtract from each raw RSS value in the FFT window the mean of a shorter window centered around this value (currently set to 5 seconds in our implementation). This \textbf{\emph{local}} mean removal balances the signal and leads to reducing the sudden change effect as shown in Figure~\ref{fig:sub_mean_removal}.

We note that using multiple RSS streams from \textbf{\emph{different heard APs}} further reduces the effect of an external user cutting the LOS with a specific AP. This is not discussed in this paper due to space constraints. In addition, the outlier detection sub-module of the \emph{Robust Breathing Rate Extractor} module described in the next section further diminishes the impact of sudden changes.

\subsection{Robust Breathing Rate Extractor}
Once both the breathing signal and the \emph{instantaneous} breathing rate are estimated, this module aims to increase the robustness of the breathing rate estimation based on fusing different overlapping consecutive windows. 
This is particulary useful in scenarios where a single stable reading is needed, e.g. to be logged in the patient's chart.  The module contains two sub-modules: the \emph{Outlier Detector} 
module and \emph{Robustness Enhancement} module. 

\subsubsection{Outlier Detector} 

To further reduce the impact of outliers (e.g. the sudden change at time $t=52$ sec in Figure~\ref{fig:dc_removal}), \sys{} uses an $\alpha$-trimmed mean filter \cite{Yang:Performin} over a sliding window of previous instantaneous estimations. An $\alpha$-trimmed mean filter sorts the values within the input window and then trims the highest and lowest $\alpha$ values. The remaining values are then averaged to obtain the smoothed filter value corresponding to the input window. More formally, given a window of $q$ sorted instantaneous breathing rate estimates ($\hat{r}_i$'s), such that $\hat{r}_1 \leq \hat{r}_2 \leq \cdots \leq \hat{r}_q$, the output of the filter $r$ is given by:
\begin{equation}
 r = \frac{1}{q- 2\left\lceil \alpha q \right\rceil} \sum_{i=\left\lceil \alpha q \right\rceil+1}^{q-\left\lceil \alpha q \right\rceil}{\hat{r}_i}
\label{eq:15}
\end{equation}
 where $0 \leq \alpha < 0.5$.

\begin{figure}[!t]
\centering
        \begin{subfigure}[t]{0.32\textwidth}
                \centering
                \includegraphics[width=\textwidth]{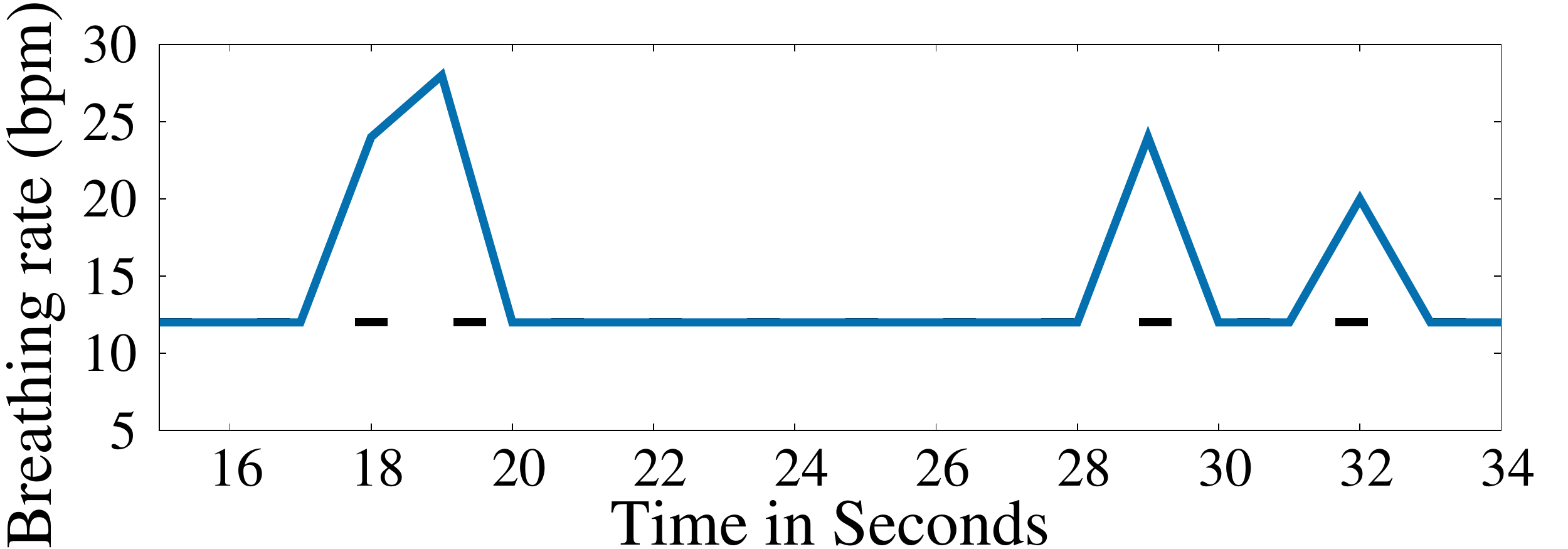}
                \caption{Instantaneous breathing rate ($\hat{r}$), the output of the \emph{Breathing Signal Extractor} module.}
                \label{fig:robustness_handling:fft_hist}
        \end{subfigure}

	\begin{subfigure}[t]{0.32\textwidth}
                \centering
                \includegraphics[width=\textwidth]{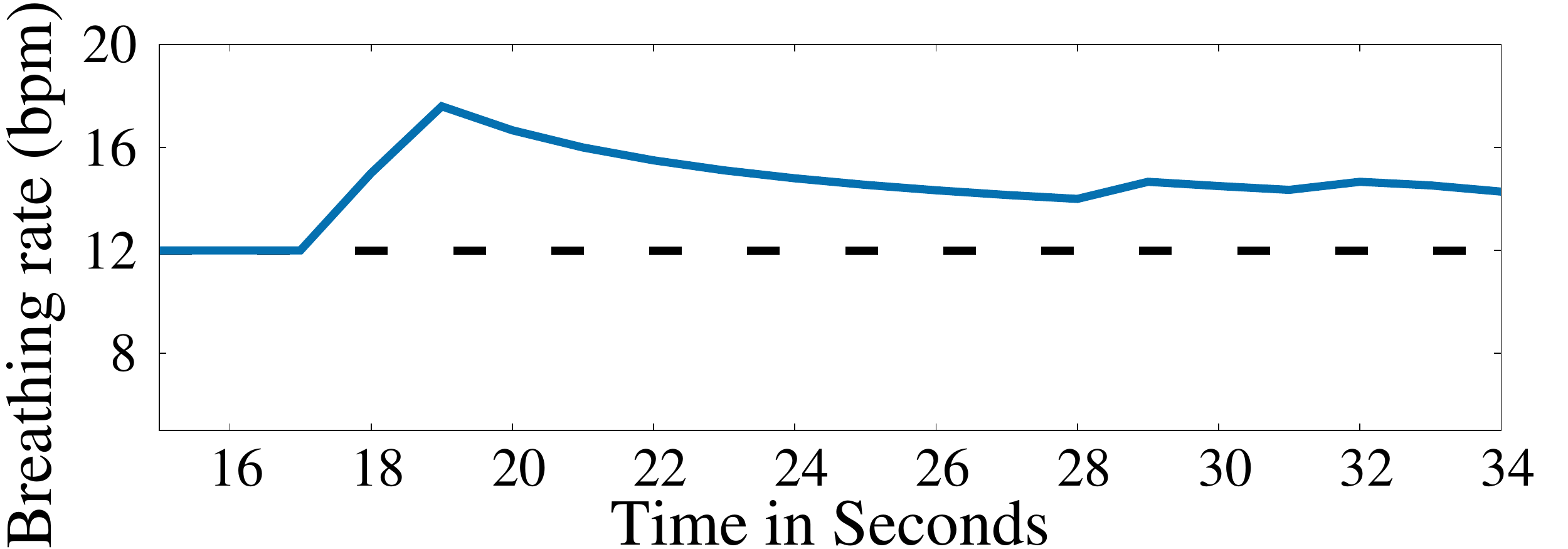}
                \caption{Output after applying a standard moving average filter.}
                \label{fig:robustness_handling:without_alpha_trimmed_filter}
        \end{subfigure}

	\begin{subfigure}[t]{0.32\textwidth}
                \centering
                \includegraphics[width=\textwidth]{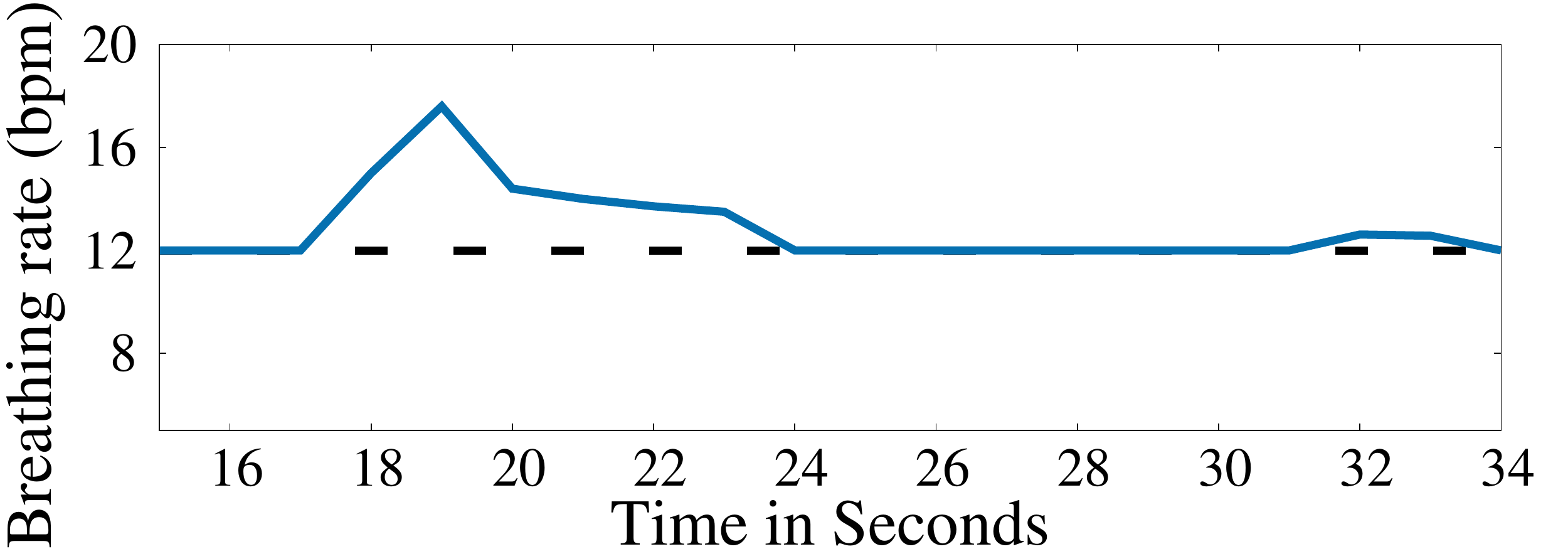}
                \caption{Output after applying a moving $\alpha$-trimmed mean filter ($\alpha = 0.25$).}
                \label{fig:robustness_handling:with_alpha_trimmed_filter}
        \end{subfigure}

\caption{Using the $\alpha$-trimmed mean filter effect for outliers detection. The black dashed line is the ground truth (12bpm). The $\alpha$-trimmed mean filter converges faster and more accurately to the true value.}
\label{fig:robustness_handling}
\vspace{-0.2in}
\end{figure}

We note that for $\alpha=0$, the filter is reduced to a standard moving average filter, while for $\alpha=0.5$, the filter is reduced to a standard median filter. An $\alpha$-trimmed mean filter has the advantage of handling both impulse and gaussian noise, as compared to median and mean filters that can handle only one of them. Therefore, we set $\alpha$ to 0.25. Figure~\ref{fig:robustness_handling} shows the effect of applying both a standard moving average filter and the $\alpha$-trimmed mean filter to the output of the \textit{Breathing Signal Extractor} module. The figure shows that the $\alpha$-trimmed mean filter can successfully reduce the effect of the outliers as well as converge faster and more accurately to the ground truth breathing rate compared to a standard moving average filter.

\subsubsection{Robustness Enhancer}
Since the breathing signal is continuously changing, \sys{} can provide a more robust breathing rate estimate compared to the instantaneous output of the $\alpha$-trimmed mean filter. The idea is to make sure that the breathing rate estimation has saturated. This is achieved by checking whether all breathing rate estimates within the last 10 seconds are consistent within a small error margin (taken as 0.75 bpm in our experiments) or not. For example, the breathing signal starts saturating after 24 seconds in Figure~\ref{fig:robustness_handling:with_alpha_trimmed_filter}. Note that the system needs to wait for the signal to saturate, introducing a slight delay for the initial robust reading. This initial delay is quantified in the evaluation section and \emph{does not affect the latency of the other modules} that operate in realtime.

\subsection{Apnea Detector}
This module aims at detecting apnea, which is the cessation of the oro-nasal airflow (i.e. the absence of breathing) for at least $10$ seconds in duration \cite{fauci2008harrison}. Apnea is the main characteristic that marks almost all abnormal breathing patterns \cite{yuanrespiratory}. During apnea, there is no movement of the inhalation muscles and the volume of the lungs remains unchanged. This fact should map to a non-changing RSS signal in \sys{}, modulo small changes caused by environmental noise. The Apnea Detector module works in two phases: denoising and apnea detection.

\begin{figure}[!t]
\centering
\includegraphics[width=3.3in]{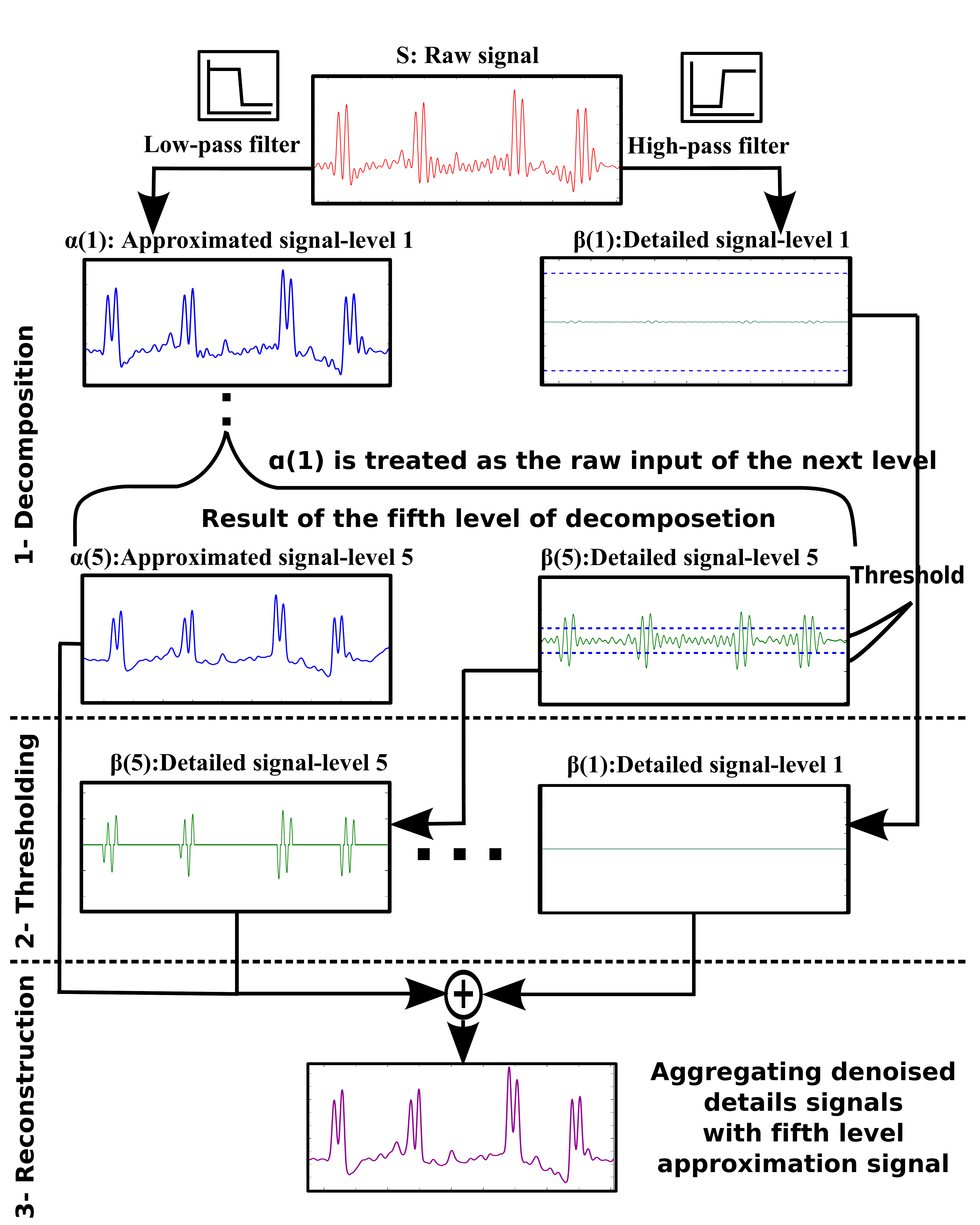}
\caption{Wavelet Denoising stages: the input signal is first decomposed using high- and low-pass filters. Then the details are thresholded using soft thresholding, such that the wavelet coefficients smaller than a given threshold (the doted blue lines) are set to zero and the coefficients above the threshold reduced by the value of the threshold. Finally, the thresholded coefficients are added to the last level approximation signal to get the reconstructed denoised signal.}
\label{fig:denoising}
\vspace{-0.1in}
\end{figure}

\subsubsection{Denoising}
To detect apnea, \sys{} further removes the noise from the breathing signal using wavelet denoising \cite{sardy2001robust}, which is based on the Discrete Wavelet Transform (DWT). In DWT, the generic step splits the signal into two parts: an approximation coefficient vector and a detail coefficient vector. This splitting is applied recursively in a number of steps (i.e. levels), $J$, to the approximation coefficient vector only to obtain finer details from the signal. At the end, DWT produces a coarse approximation projection (scaling) coefficients $\alpha^{(J)}$, together with a sequence of finer detail projection (wavelet) coefficients $\beta^{(1)},  \beta^{(2)}, ...,\beta^{(J)}$. The DWT coefficients in each level can be computed using the following equations:
\begin{equation}
\alpha^{(J)}_k = \langle x_n, g^{(J)}_{n-2^J k} \rangle_n = \sum_{n \in \mathbb{Z} } x_n ~g^{(J)}_{n-2^J k},	~~~~	J \in \mathbb{Z}
\label{eq:approximations}
\end{equation}
\begin{equation}
\beta^{(\ell)}_k = \langle x_n, h^{(\ell)}_{n-2^l k} \rangle_n = \sum_{n \in \mathbb{Z} } x_n ~h^{(\ell)}_{n-2^\ell k},	~~~~	\ell \in \{1, 2, ..., J\}
\label{eq:details}
\end{equation}
where $x_n$ is the $n^\textrm{th}$ input point, $\langle . \rangle$ is the dot product operation, and $g$'s and $h$'s are two sets of discrete orthogonal functions called the wavelet basis (we used the Haar basis functions in our system). 
 The inverse DWT is given by
\begin{equation}
x_n = \sum_{k \in \mathbb{Z} } \alpha^{(J)}_k ~g^{(J)}_{n-2^J k} + \sum^J_{\ell=1} \sum_{k \in \mathbb{Z} }\beta^{(\ell)}_k ~h^{(\ell)}_{n-2^\ell k}
\label{eq:idwt}
\end{equation}

Wavelet denoising builds on the DWT as shown in Figure~\ref{fig:denoising} by running through three phases. In the beginning, the raw input signal is subjected to a decomposition phase using DWT to break the signal down into an approximation signal, and a wavelet detail coefficient. This process is conducted recursively to each resulting approximation signal up to five levels. Afterwards, dynamic thresholding is applied to the wavelet detail coefficients resulting from each level to remove their noisy components. Finally, inverse wavelet transform is performed on all the resulting coefficients, in addition to the approximated signal from the last level, to reconstruct a final denoised signal. Compared to other techniques, wavelet denoising runs efficiently in linear time with respect to the size of the input signal. In addition, it does not make any assumptions about the nature of the signal and permits discontinuities in the input signal \cite{nibhanupudi2003signal} (similar to those caused by apnea). 

\begin{figure}[!t]
\centering
	\begin{subfigure}[t]{1.645in}
                \centering
                \begin{minipage}{\linewidth}
                		\includegraphics[width=\textwidth,height=1.2cm]{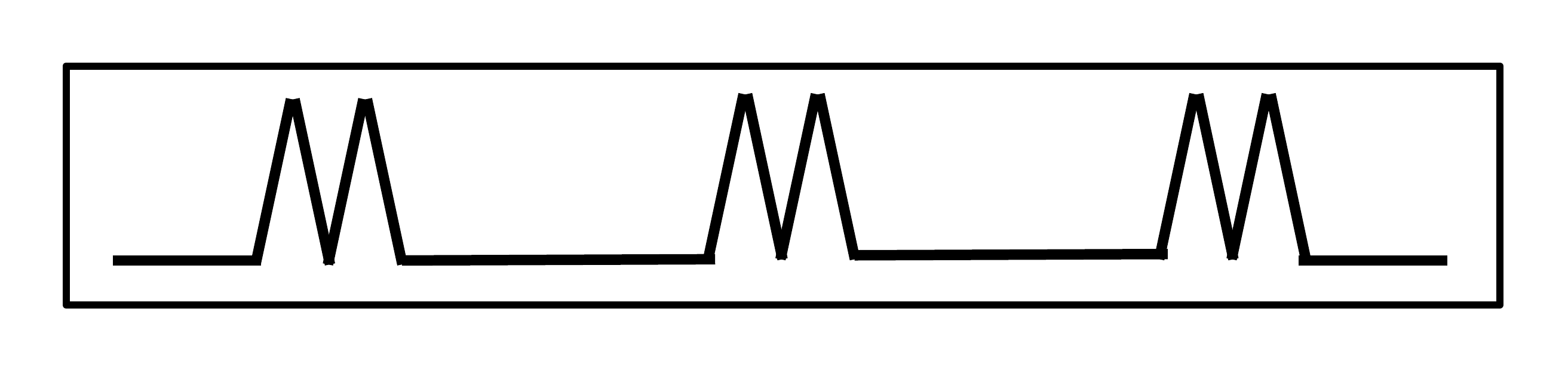}
                		\caption*{}

               		\includegraphics[width=\textwidth,height=1cm]{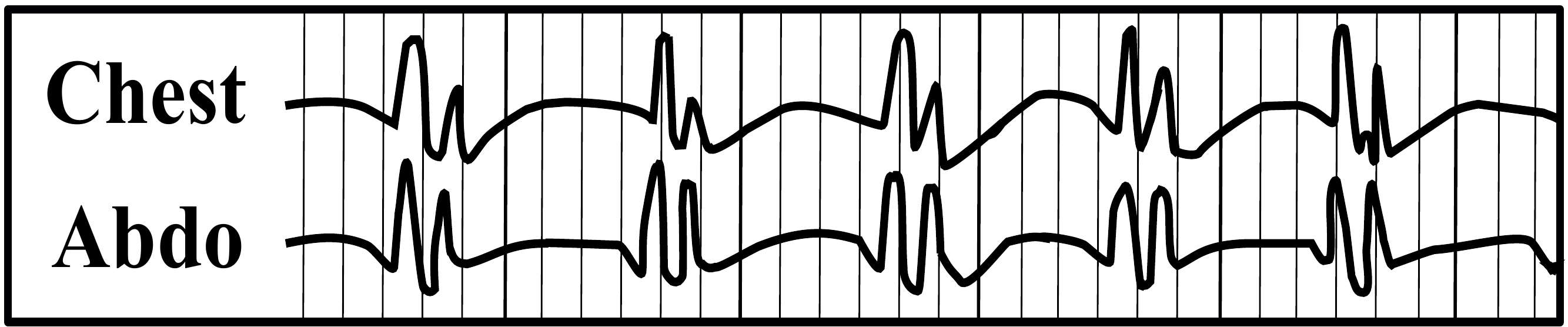}
                		\caption*{}

                		\includegraphics[width=\textwidth]{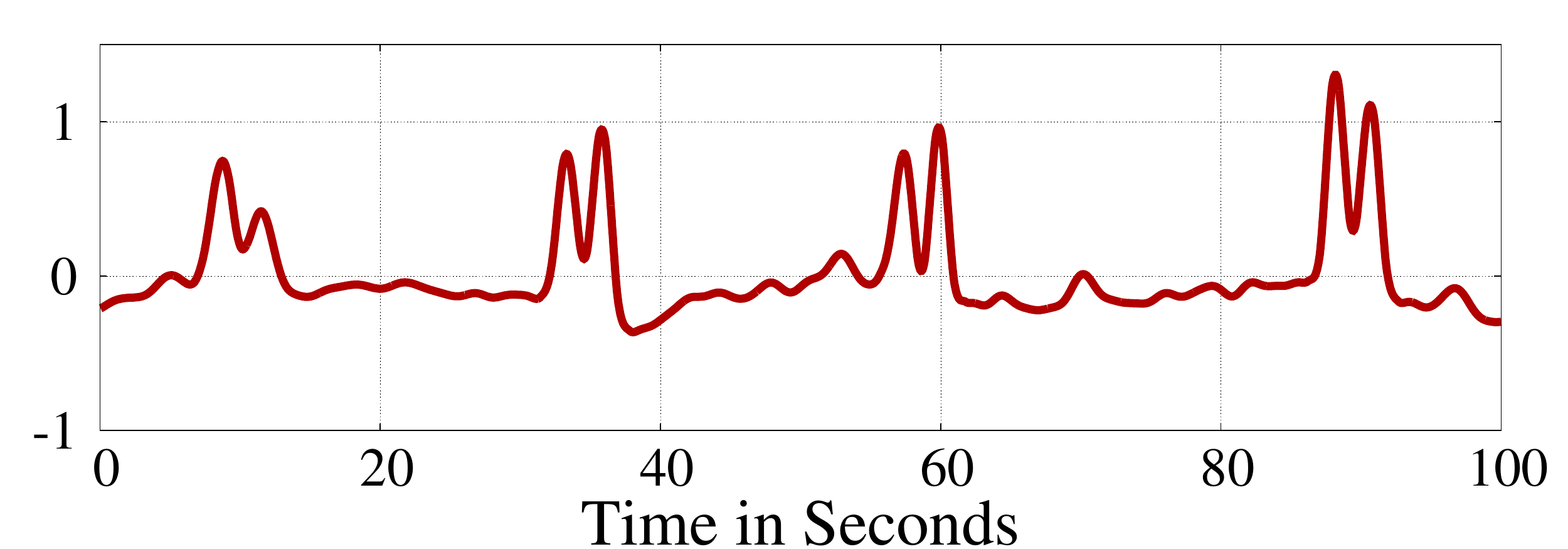}
                		\caption*{}
                \end{minipage}
                \caption{Biot's Respiration: a respiratory pattern characterized by periods or ``clusters'' of rapid respirations of near equal depth followed by regular periods of apnea.}
	      \label{fig:boits}
        \end{subfigure}
        \begin{subfigure}[t]{1.645in}
                \centering
                \begin{minipage}{\linewidth}
                		\includegraphics[width=\textwidth,height=1.2cm]{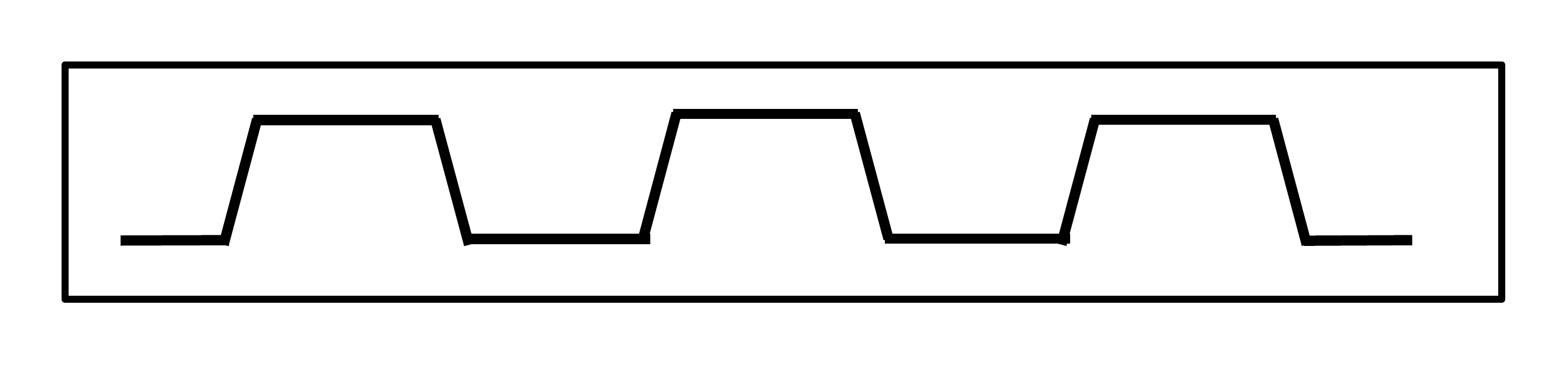}
                		\caption*{\hspace{-4.7cm}Ground truth pattern.}

                		\includegraphics[width=\textwidth,height=0.95cm]{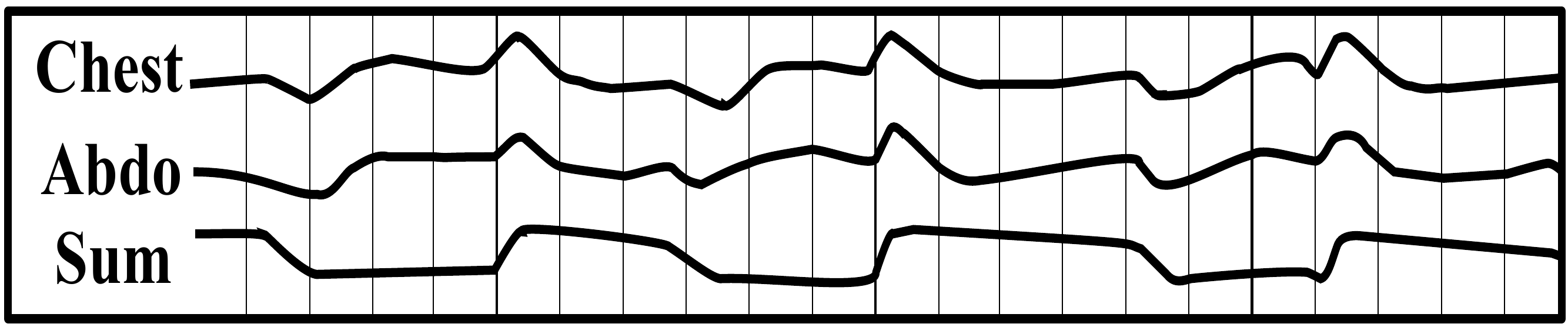}
                		\caption*{\hspace{-4.25cm}\mbox{Measured by respiratory inductance plethysmography \cite{yuanrespiratory}.}}

                		\includegraphics[width=\textwidth]{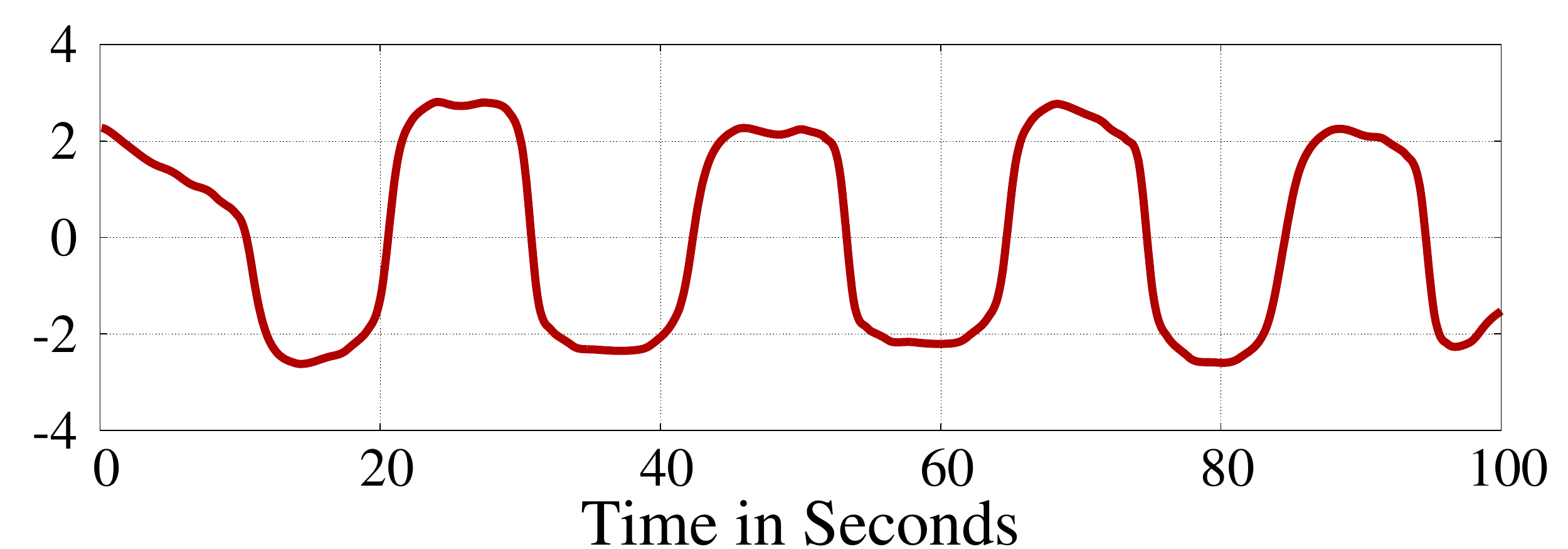}
                		\caption*{\hspace{-4.5cm}Measured by \sys{}.}
               \end{minipage}
               \caption{Apneustic Respiration: has a prolonged inspiratory phase followed by a prolonged expiratory phase commonly believed to be apneic phases \cite{yuanrespiratory}.}
	      \label{fig:apneustic}
        \end{subfigure}
        \caption{The figures compare between the visualization of respiratory inductance plethysmography that typically uses invasive techniques and the visualization of \sys{}.}
        \label{fig:abnormal_patterns}
\vspace{-0.1in}
\end{figure}
\newpage
\subsubsection{Apnea detection}

Some of the typical apnea patterns, involving repeated prolonged loss of breath, are shown in Figure~\ref{fig:abnormal_patterns}. To detect the loss of the breathing signal for more than $10$ seconds, a sliding moving window of samples covering 10 seconds is used over the denoised input signal. The difference between the maximum and minimum signal values within this window is compared to a threshold ($\theta$). If this difference is below the threshold, then there is no breathing signal and an apnea alarm is raised. More formally, for a given window of denoised samples $y[1..m]$, an alarm is raised if:

\begin{equation}
  \max(y[1..m])-\min(y[1..m])\, \substack{\textrm{Apnea Det.}\\\lessgtr\\\textrm{No Apnea}}\, \theta
\end{equation}
where $\theta$ is the detection threshold.

To make apnea detection adapt to dynamic environments, we use a dynamic threshold rather than a fixed one. The threshold ($\theta$) is taken as a percentage ($\tau$) of the range of the last detected breathing signal. We quantify the effect of $\tau$ on performance in the evaluation section. Overall, Figure~\ref{fig:abnormal_patterns} compares the results of \sys{} when applied to a person emulating the ground truth patterns shown at the top of the figure, with the plethysmography measurements of those with the same apnea pattern.

Finally, \sys{} leverages multiple RSS streams received from the different APs naturally deployed in an environment to increase the system accuracy and resilience to noise. In this case, a simple majority vote on the detected state (apnea/ normal breathing) from each stream is used to fuse the detection of the different APs for enhanced accuracy.

\subsection{Realtime Visualization Module}

The goal of the \emph{Realtime Visualization} module is to combine the output of the different system modules in a user friendly manner. In particular, the output from these modules is collected and streamed to a pre-defined user device, e.g. a laptop to show the breathing signal, the breathing rate, as well as raise an audible and visual alarm if apnea is detected.

Among these outputs, the breathing signal is of specific importance. Since abnormal respiratory rates and changes in them are broad indicators of major physiological instability \cite{knaus1985apache}, displaying the breathing waveform can help medical practitioners detect different breathing anomalies.
In particular, normal ventilation is an automatic, seemingly effortless inspiratory expansion and expiratory contraction of the chest cage. This act of normal breathing has a relatively constant rate and inspiratory volume that together constitute the normal respiratory rhythm (inspiration = expiration). Abnormality may occur in the rate, rhythm, and in the effort of breathing. These properties are clear in the RSS waveform after being processed by \sys{} as shown in Figure~\ref{fig:abnormal_patterns}. Note that, due to space constraints, we only provide the details for detecting apnea and leave detecting its specific type to future work.

\subsection{Discussion}
\label{sec:discuss}
\sys{} applies a number of techniques to handle the noisy wireless environment and interfering nearby users. Specifically, using a band pass filter within the normal human breathing rates helps in removing other sources of interference in the frequency spectrum. Moreover, \sys{} explicitly handles sudden changes in the received WiFi signal using the local mean removal technique. In addition, using the $\alpha$-trimmed mean filter helps in removing outliers and speeds the convergence time.

The system can leverage multiple APs to increase the accuracy of both the breathing rate estimation and apnea detection as well as reduce the effect of interfering humans that usually interfere with the stream from only one AP.

The design of the user interface that initiates the estimation process based on user request makes the system more resilient to outliers and helps in reducing its computational requirements. In addition, since the measurements are done at the user's device, multiple users can estimate their own breathing rate in \textbf{\emph{parallel}}, with no cross-interference on each other leveraging the broadcast nature of the wireless channel.

Since human breathing rate is a low frequency signal, the normal beacon transmission rate of WiFi (IEEE 802.11 standard) is well above the Nyquist frequency. This makes \sys{}'s energy consumption within the normal range required for WiFi operation.

\section{Evaluation}
\label{evaluation}

\begin{figure}[!b]
\vspace{-0.1in}
\centering
	\begin{subfigure}[b]{0.4\textwidth}
	\includegraphics[width=\textwidth]{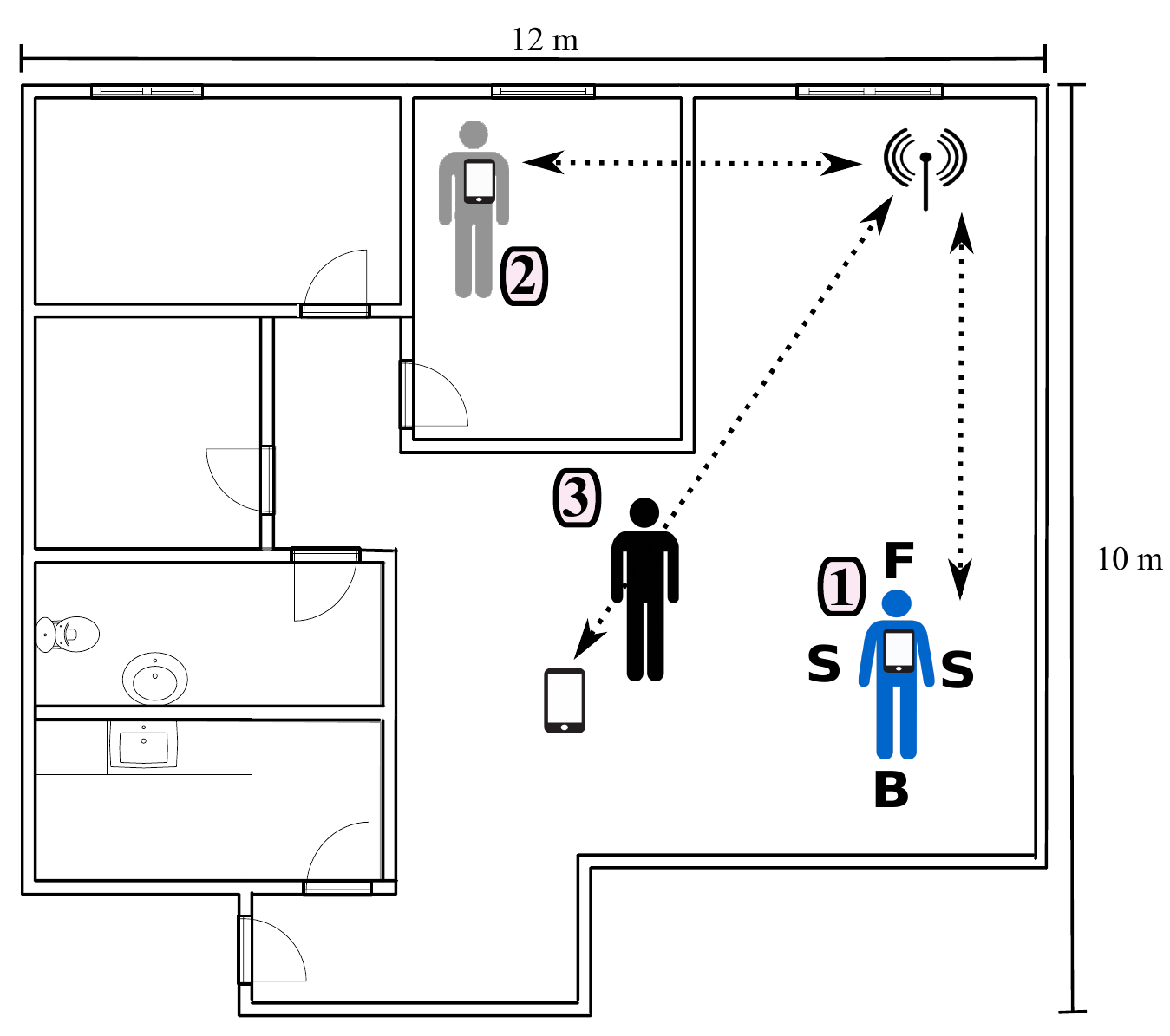}
	\caption{First test environment (apartment).}
	\label{fig:apartment_floorplan}
	\end{subfigure}
\centering
	\begin{subfigure}[b]{0.45\textwidth}
	\includegraphics[width=\textwidth]{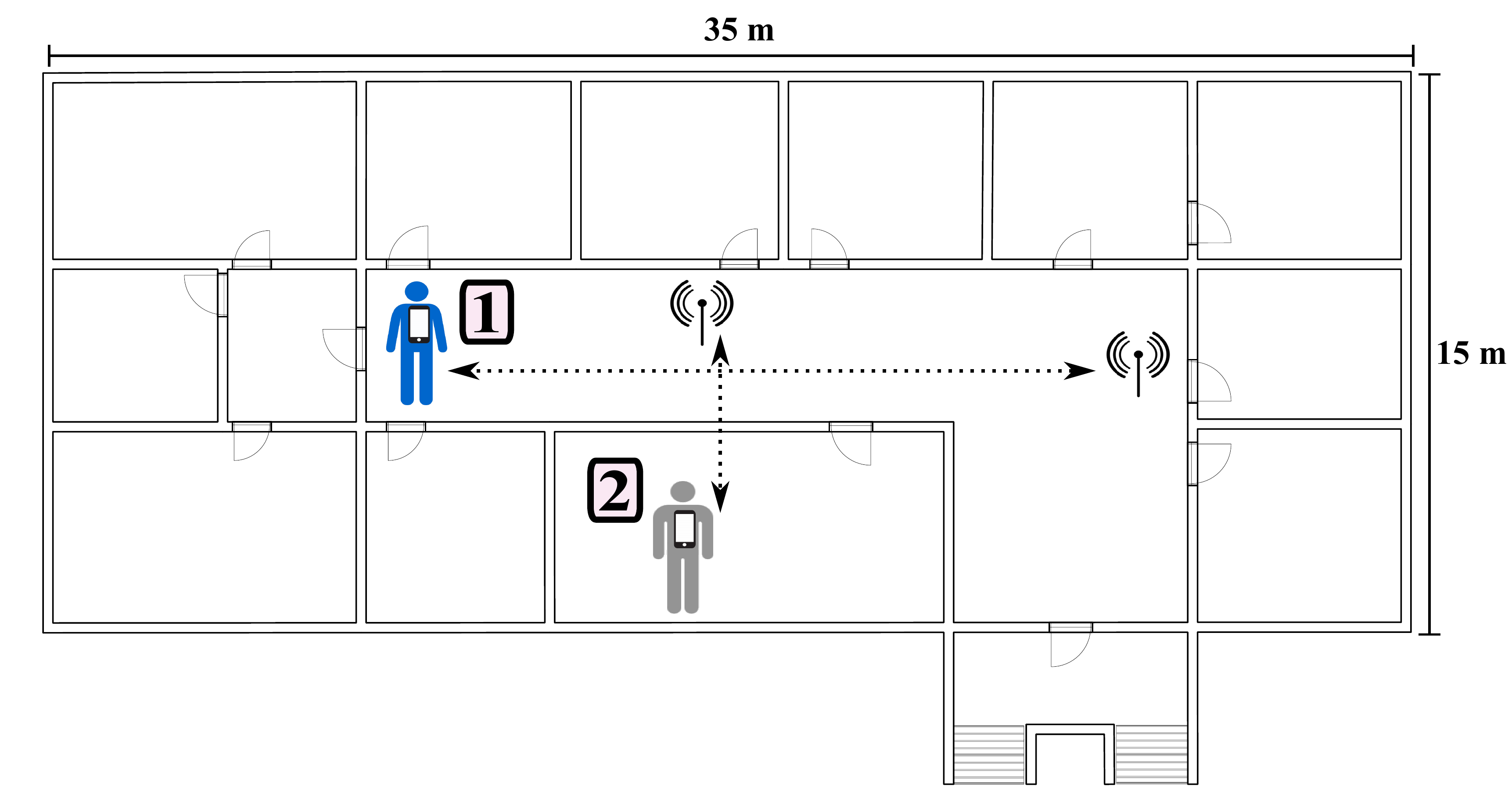}
	\caption{Second test environment (one floor of the engineering building where our lab is located).}
	\label{fig:vtmena_floorplan}
	\end{subfigure}
\caption{Floorplans of the two test environments used for evaluating \sys{}. The labeled persons denote different \sys{} scenarios, 1) Holding-the-device: The user is holding the mobile device on her chest. 2) Holding-the-device through-the-wall: The user and AP are in two rooms while the user is holding the mobile device on her chest. 3) Hands-free On LOS: The user is standing between the AP and the device. The orientation characters in holding-the-device scenarios indicate the possible user orientations relative to the AP.}
\label{fig:testbeds}
\end{figure}

In this section, we analyze the performance of \sys{} in two typical environments as depicted in Figure~\ref{fig:testbeds}: (a) An apartment covering a 10$\times$12 $\textrm{m}^2$ area and consisting of a living room connected to the dining area, three bedrooms, kitchen, and a bathroom. (b) The second floor of our engineering building which contains 12 rooms connected by a large corridor with a total area of 35$\times$15 $\textrm{m}^2$.

In both environments, a Samsung Galaxy S4 Mini mobile and a Samsung Galaxy Note II devices as well as an HP EliteBook laptop are used as receivers with the typical WiFi sampling rate of $10$Hz. The used APs are Cisco Linksys X2000. We run the experiments with three different users at different times of the day over a period of two weeks. A custom-made breathing metronome application that instructs the users about their breathing rate/time is used to regulate the breathing rate of these users and acts as the ground truth. In all experiments, there were a varying number of interfering users in their normal daily activities, reaching up to three users in the same room, in addition to other unknown persons in the other rooms and offices. 

We extensively evaluate \sys{} in three different scenarios shown in Figure~\ref{fig:testbeds}: \textbf{1) Holding-the-device}: The user is holding the mobile device on her chest, \textbf{2) Holding-the-device, through-the-wall}: The user and AP are in two different rooms while the user is holding the mobile device on her chest, and \textbf{3) Hands-free On-LOS}: The user is standing between the AP and the device. We evaluate the system accuracy in these three scenarios and then evaluate it for multiple users breathing with different rates. Table~\ref{tab:def_parameters} shows the default values of the system parameters utilized in our experiments. All accuracy numbers are based on the output of the \textit{Robust Breathing Rate Extractor} module.

\begin{table}[!t]
    \centering
    \begin{tabular}{|p{3cm}|c|c|} \hline
      \textbf{Parameter} & \textbf{Range} & \textbf{Default value} \\ \hline \hline
      FFT window  size ($w$)& $10-60$ $\textrm{sec}$ & $30 \textrm{sec}$ \\ \hline
      Sampling rate (Fs) & $1-10$ Hz & $10$ Hz  \\ \hline
      Breathing rate ground truth& $12,18, 24$ $\textrm{bpm}$ & $18 \textrm{bpm}$ \\ \hline
      Distance between the AP and mobile device& $1-14$ $\textrm{m}$ & $8$ $\textrm{m}$  \\ \hline
       Apnea detection threshold ($\tau$) & $33$ - $67\%$ & $50\%$ , $67\%$ \\ \hline
       Number of Streams & 1-5 & 1 \\ \hline
      Orientation & Front, side, back & Front \\ \hline 
      Testbeds & Apt., Eng. building& Both \\
      \hline
  \end{tabular}
    \caption{Parameter ranges and default values.}
    \label{tab:def_parameters}
    \vspace{-0.2in}
\end{table}

\subsection{Holding the Device Scenario}
In this section, we study the system performance in the holding-the-device scenario under different sampling rates, distances from the AP, and user orientations. Both the LOS and through-the-wall scenarios are evaluated.

\subsubsection{Impact of sampling parameters}
\begin{figure}[!t]
\centering
        \begin{subfigure}[t]{0.35\textwidth} 
                \centering
                \includegraphics[width=\textwidth, height=1.6in]{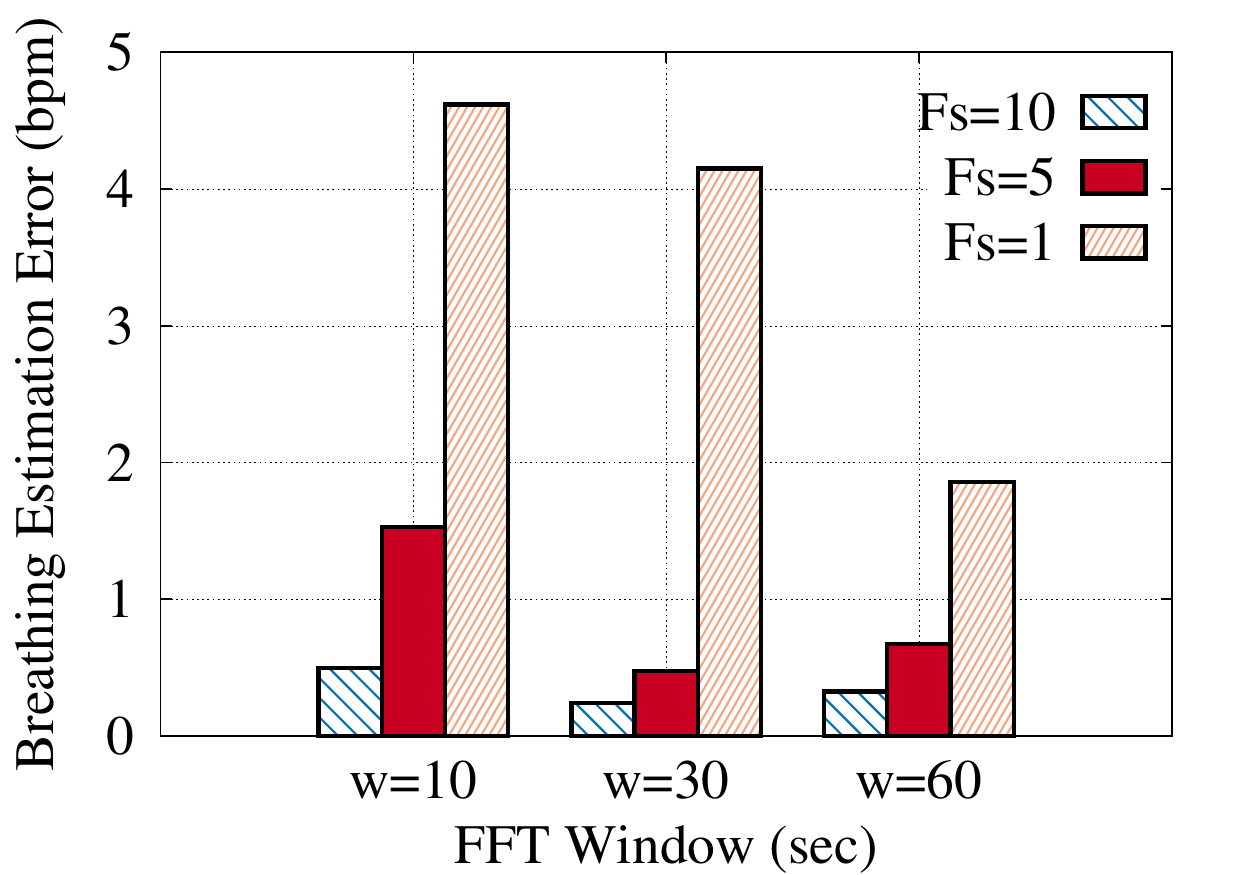}
                \caption{Estimation error.}
                \label{fig:sampling_error}
        \end{subfigure}
	\begin{subfigure}[t]{0.34\textwidth} 
                \centering
                \includegraphics[width=\textwidth, height=1.6in]{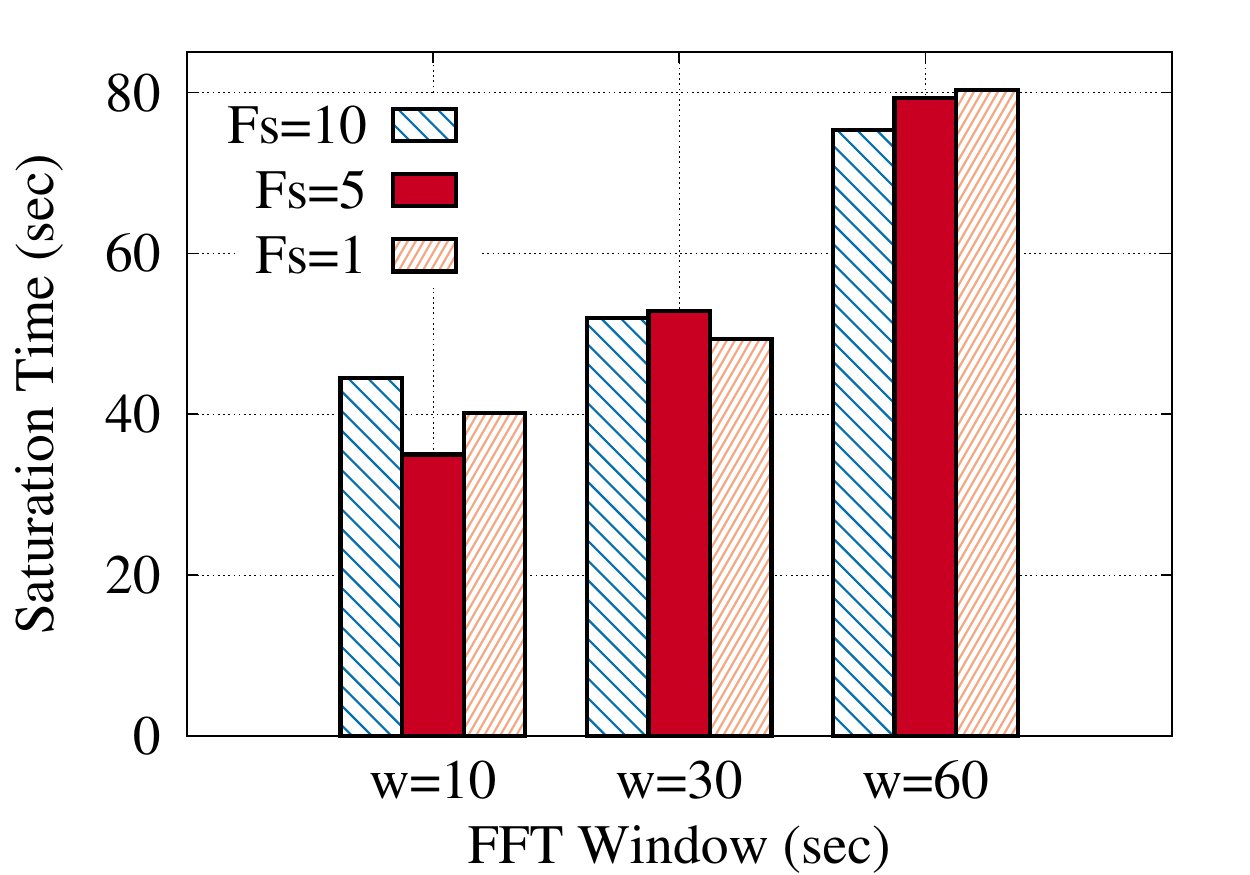}
                \caption{Required \textbf{initial} time to saturate (latency of the \emph{Robustness Enhancer} module). Other modules work in \textbf{\emph{realtime}}.}
                \label{fig:sampling_time}
        \end{subfigure}
\caption{Effect of sampling parameters on system performance for the device-on-chest scenario. The ground truth breathing rate is 18 bpm. Other parameters are set to their default values in Table~\ref{tab:def_parameters}.}
\label{fig:sampling_rate}
\vspace{-0.1in}
\end{figure}

Figure~\ref{fig:sampling_rate} shows the system performance for different sampling rates ($\textrm{Fs}$) and different FFT window sizes (parameter $w$). \textbf{\emph{Other parameters are set to their default values as in Table~\ref{tab:def_parameters}.}} The figure shows that, in general, the accuracy of breathing estimation increases with the increase in the FFT window size. However, this accuracy comes at the cost of increasing the \textbf{\emph{initial}} latency of the \emph{Robustness Enhancer} module. Using the typical WiFi beaconing rate of 10 Hz, we can obtain a low estimation error of $0.2$ bpm using a window size of $30$ sec. We use these values for the remainder of our evaluation section.

\subsubsection{Impact of distance separating the device and AP}

\begin{figure}[!t]
\centering
	\begin{subfigure}[t]{0.23\textwidth} 
                \centering
                \includegraphics[width=\textwidth]{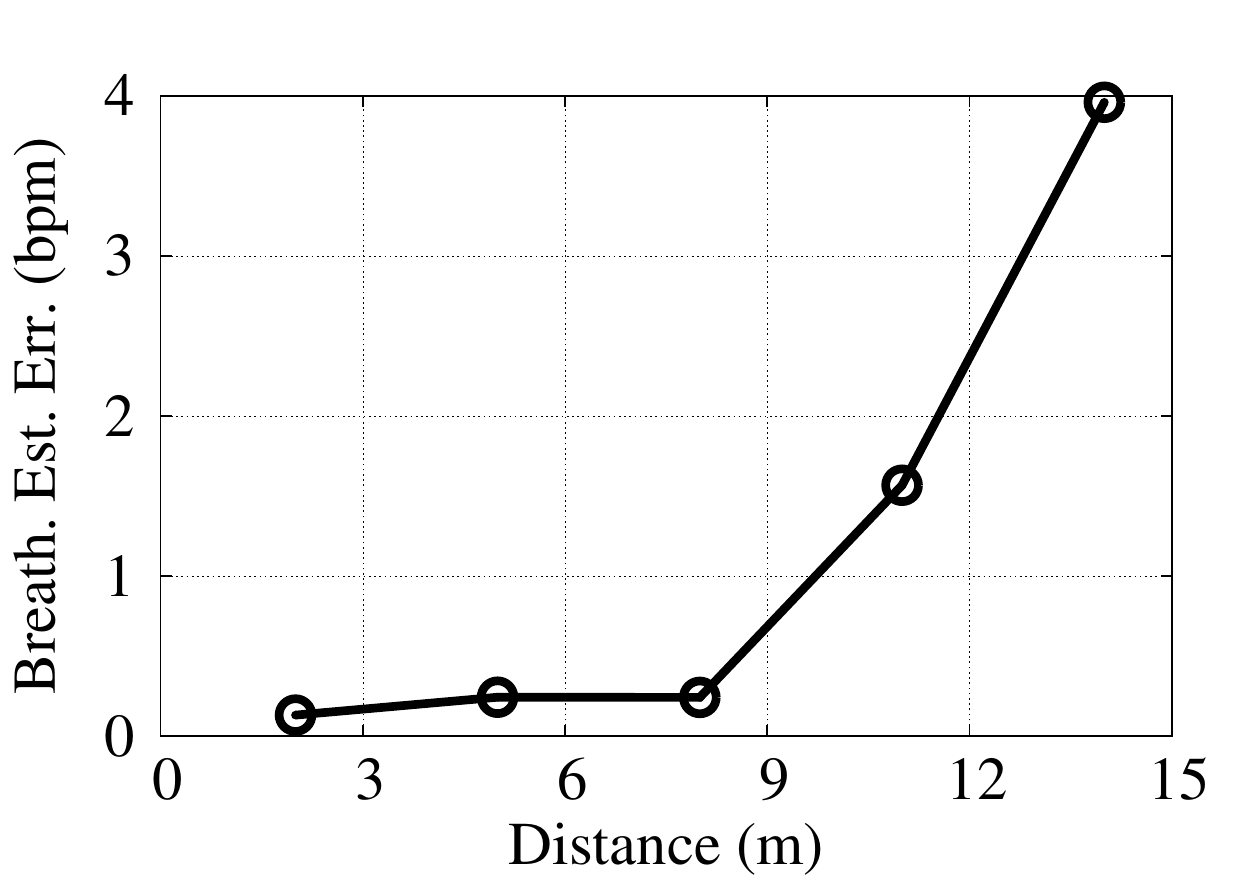}
                \caption{No-wall.}
                \label{fig:no_wall}
        \end{subfigure}
	\begin{subfigure}[t]{0.23\textwidth} 
                \centering
                \includegraphics[width=\textwidth]{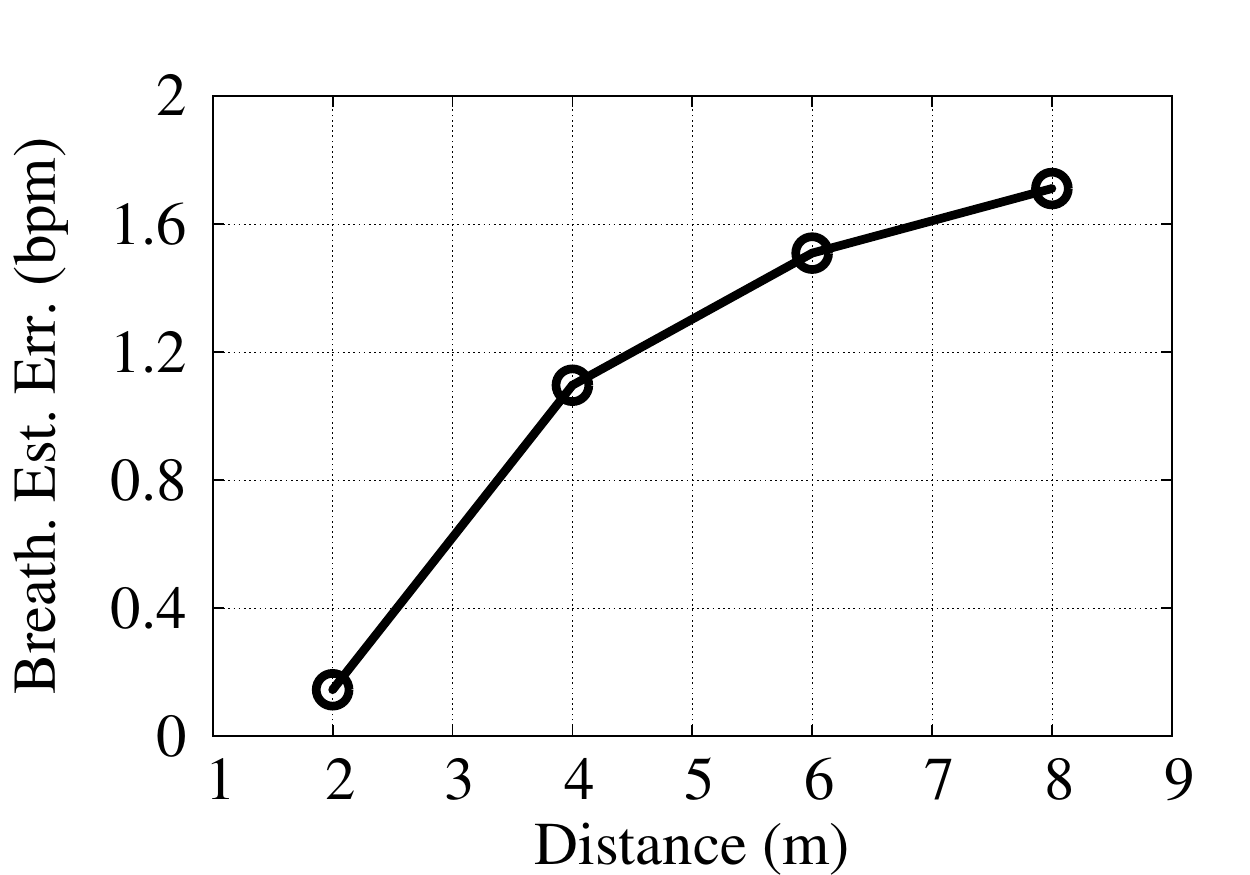}
                \caption{Through-the-wall.}
                \label{fig:through_wall}
        \end{subfigure}
\caption{Effect of the distance between the AP and user device on system performance for the device-on-chest scenario.}
\label{fig:distance}
\vspace{-0.1in}
\end{figure}

Figure~\ref{fig:distance} shows the effect of changing the distance between the device and the AP on the system's accuracy. We observe that the breathing rate estimation accuracy decreases when increasing the separation distance. This is because a stronger RSS leads to a higher signal-to-noise ratio (SNR) and hence better accuracy. In other words, a weaker signal leads to lower changes in the signal strength in response to the chest movement, leading to less sensitivity. \sys{}, however, can still be within $1.6$bpm from the ground truth for distances up to $11 \textrm{m}$ in the no-wall scenario and $1.7$bpm for distances up to $8 \textrm{m}$ in through-the-wall scenario.  

\begin{figure}[!t]
\centering
\includegraphics[width=2.5in]{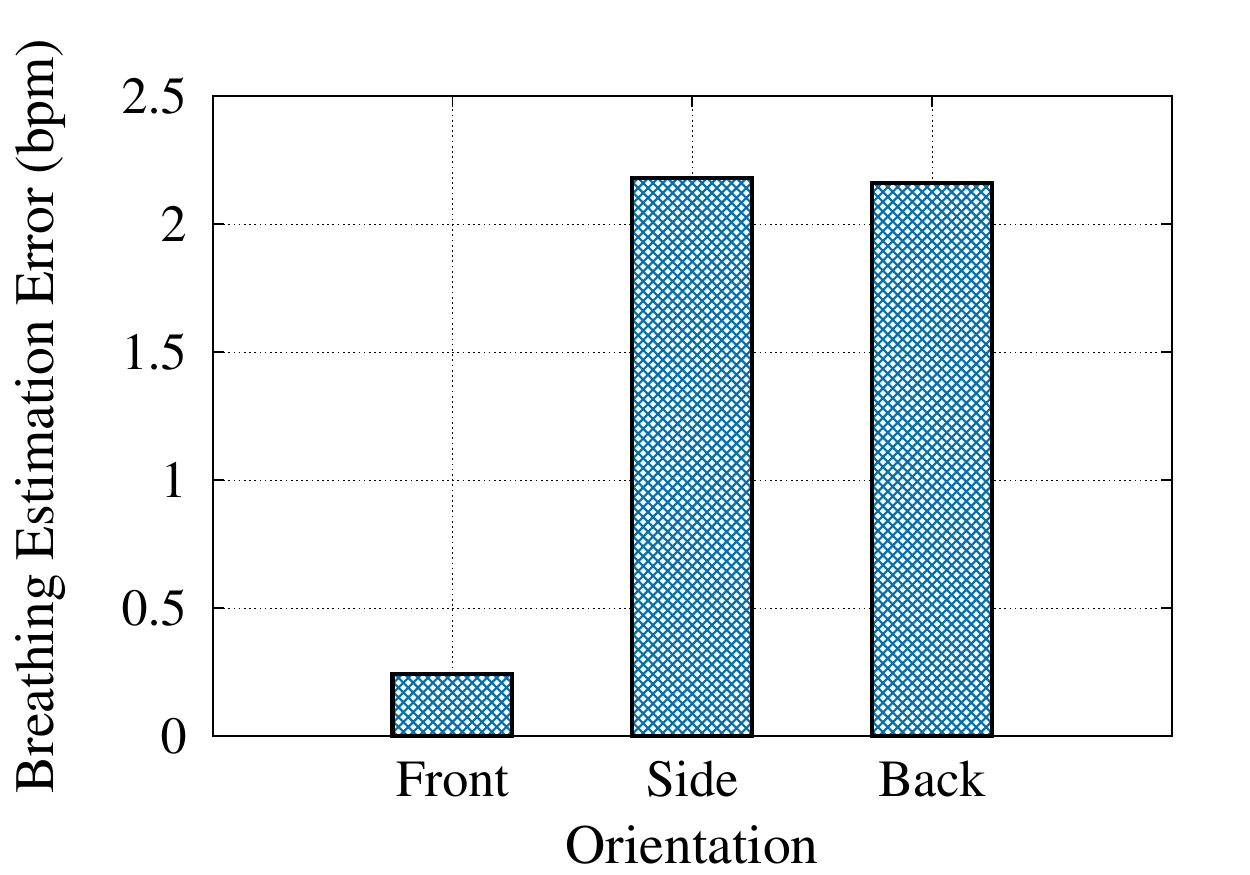} 
\caption{Effect of the user orientation relative to the AP on performance for the device-on-chest scenario.}
\label{fig:orientation}
\vspace{-0.1in}
\end{figure}

\begin{figure*}[!t]
\centering
	\begin{subfigure}[t]{0.33\textwidth} 
        \centering
        \includegraphics[width=2.4in, height=1.7in]{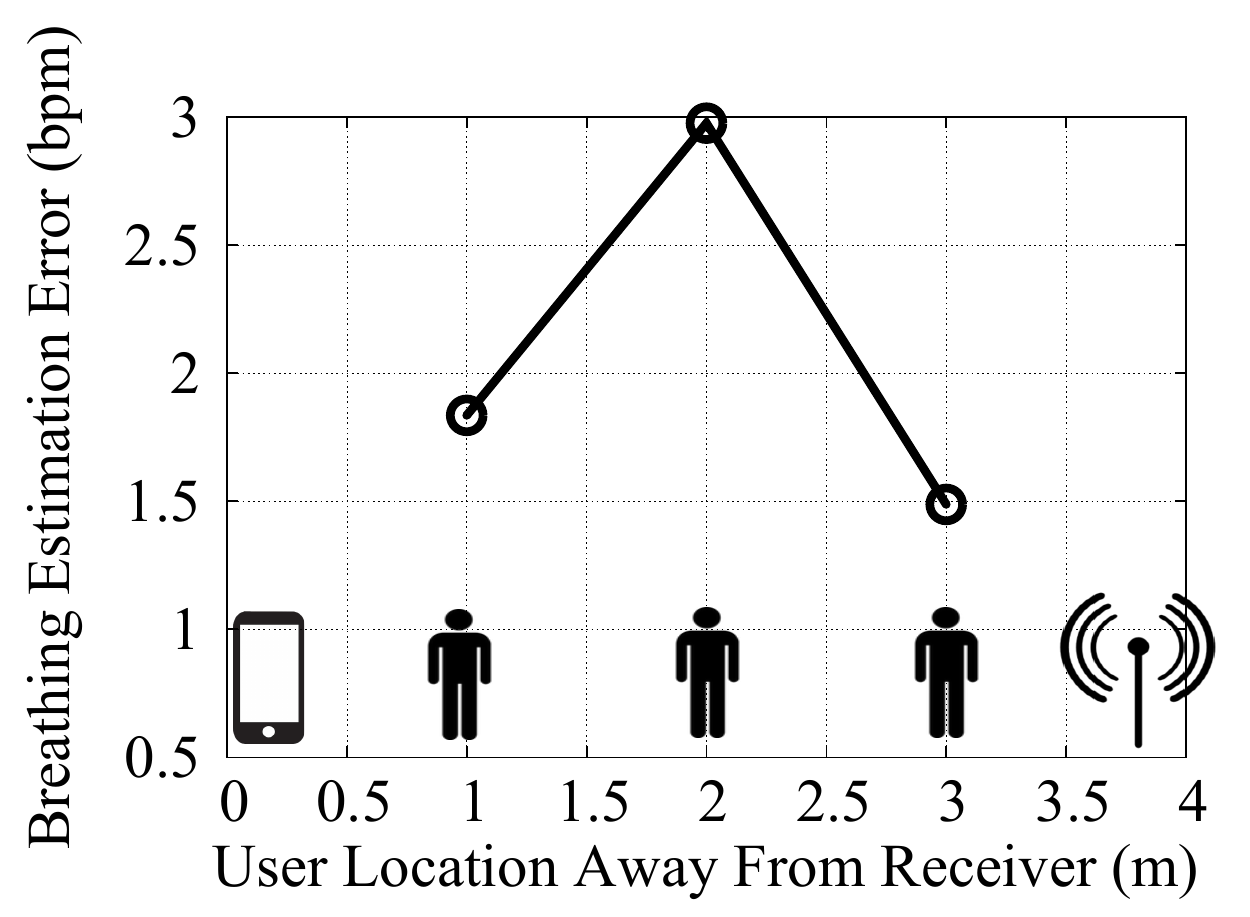}
        \caption{The user at different positions between the AP and mobile device.}
        \label{fig:person_distance_relativeto_nodes}
    \end{subfigure}
	\begin{subfigure}[t]{0.33\textwidth} 
        \centering
        \includegraphics[width=2.4in, height=1.7in]{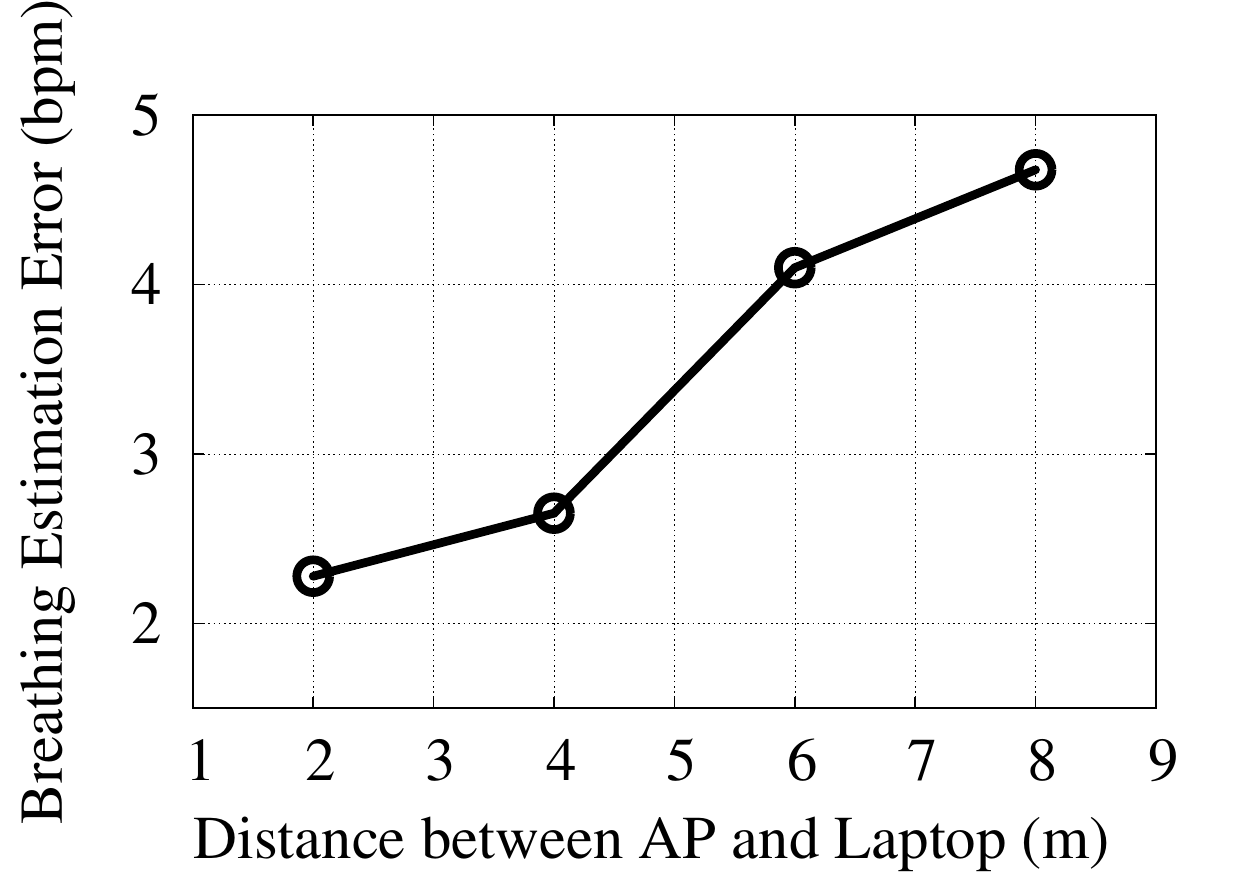}
        \caption{Varying the distance between the AP and device with user fixed.}
        \label{fig:distance_bet_APandLaptop}
    \end{subfigure}
    \begin{subfigure}[t]{0.33\textwidth} 
        \centering
        \includegraphics[width=2.4in, height=1.7in]{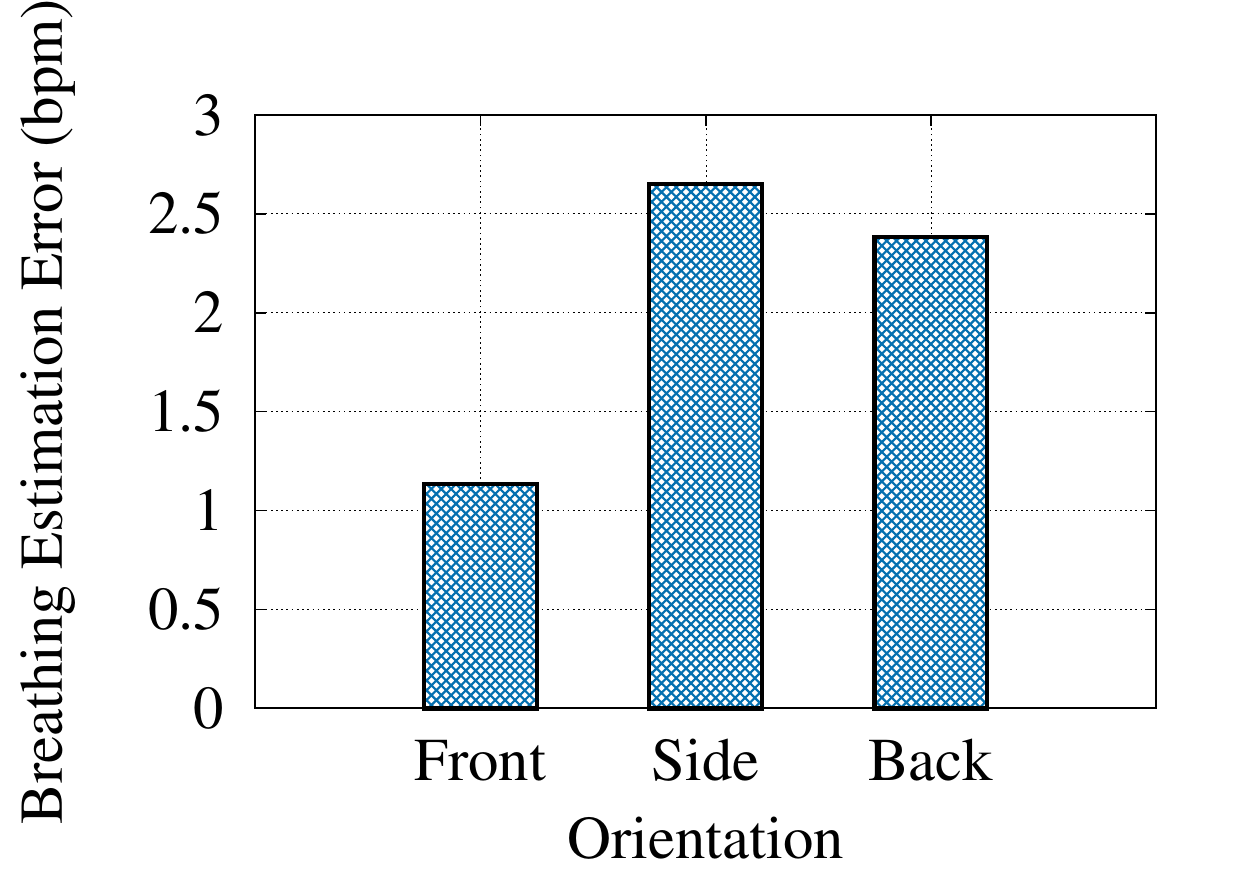}
        \caption{The impact of user orientation relative to the AP.}
        \label{fig:los_orientation}
    \end{subfigure}
\caption{This figure shows the performance of \sys{} in terms of breathing estimation error under different hands-free scenarios including: (a) varying the position of the user between the AP and device, (b) varying the distance between the AP and device while keeping the user fixed $1m$ away from the mobile device, and (c) examining the performance with different user orientations.}
\vspace{-0.1in}
\end{figure*}

\subsubsection{Impact of user orientation relative to the AP}
Figure~\ref{fig:orientation} shows the effect of different user orientations relative to the AP (front, back, and side) on the accuracy of \sys{}. We observe that the system achieves the highest performance (estimation error of $0.24$bpm) in the front orientation. This result occurs when the user's body is essentially not blocking the LOS, and hence, the received signal is relatively stronger compared to the other orientations. For the other two orientations (side and back), \sys{} still maintains a high estimation accuracy of less than $2.5$bpm error.

\subsection{Hands-free in LOS Scenario}
We now move to the in-LOS scenario (i.e. hands-free scenario) shown in Figure~\ref{fig:testbeds} where the user is not carrying the device but rather standing between the AP and device. This scenario is suitable for many applications that require some form of constant monitoring during long periods of time, such as infant or elderly breathing monitoring during their sleep.

\subsubsection{Impact of the user distance relative to the device and AP}

In this experiment, the distance between the AP and the device is set to $4 \textrm{m}$, and the error rate is recorded for users situated at different points within this range. Figure~\ref{fig:person_distance_relativeto_nodes} shows the results. We can observe that the system's performance is improved when the user is standing near the AP or the mobile device than when she is standing in the middle. This result is due to the fact that when the user is closer to either the transmitter or receiver, she blocks the majority of signal paths between them. Given the application scenarios mentioned earlier, it is highly likely to have users closer to one of the devices (e.g. when a person has a mobile device next to her during sleep).

\subsubsection{Impact of distance between the device and AP}
In this experiment, we evaluate the system performance for different distances between the AP and device in the hands-free scenario. The purpose of this experiment is to obtain an idea of when the system starts becoming less accurate relative to typical room sizes where such monitoring would take place. The user is stationary at a distance of $1$ meter away from the mobile device. Figure~\ref{fig:distance_bet_APandLaptop} shows that, similar to the mobile on chest case, the system performance decreases when the distance between the AP and the mobile device increases due to the noisier signal. 

\subsubsection{Impact of orientation}
Figure~\ref{fig:los_orientation} shows the effect of different user orientations relative to the AP. The figure shows that, similar to the device on chest case, the system achieves the highest performance in the front orientation, with a worst case accuracy among all orientations of $2.6$ bpm error. 

\subsection{Multiple Persons Case Study}
\begin{figure}[!t]
\centering
  \includegraphics[width=2.8in]{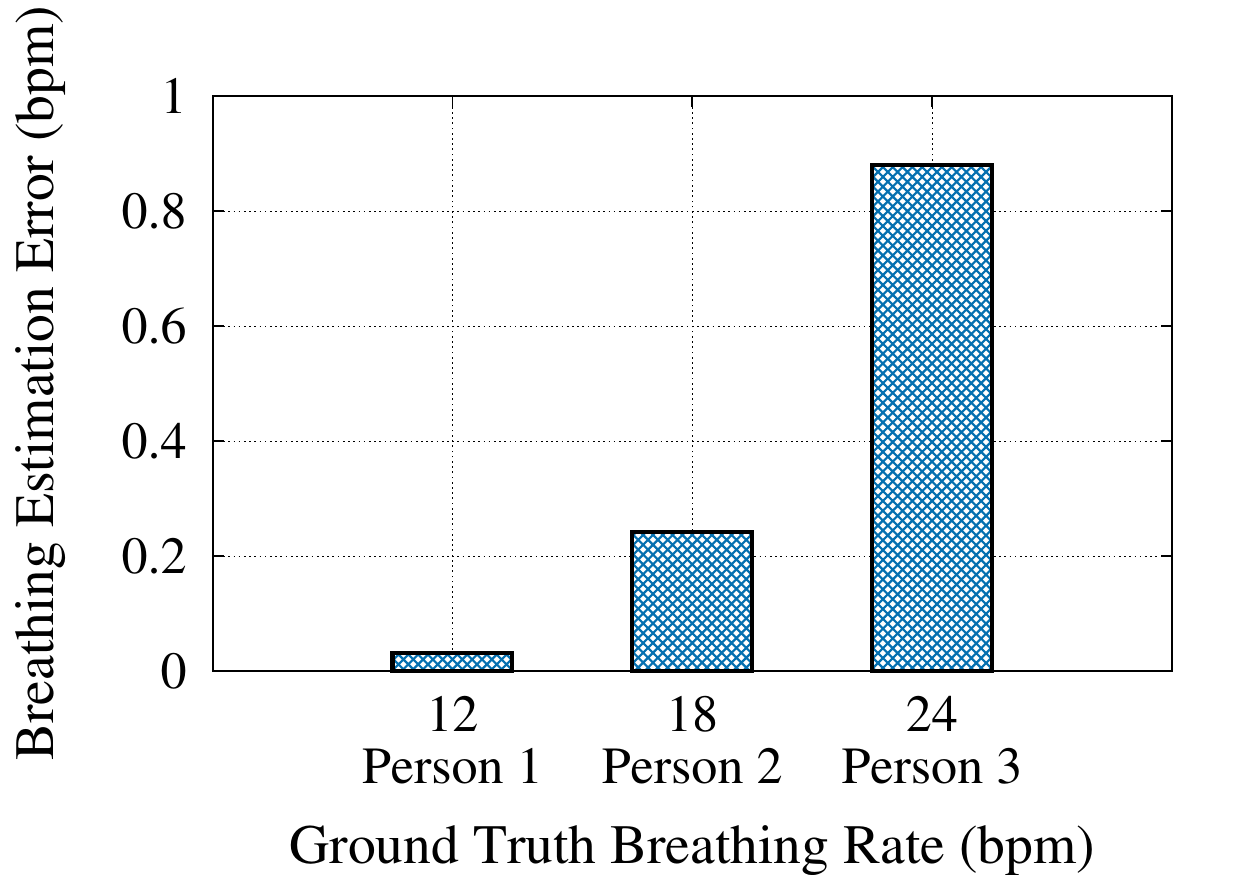}
  \caption{Performance for three concurrent measurements from different persons with different breathing rates.}\label{fig:breathing_rates}
\end{figure}

We evaluate the system in the apartment environment using one AP and three users \textbf{\emph{concurrently}} holding their devices while breathing with different rates $12$, $18$, and $24$ bpm (as in Figure~\ref{fig:apartment_floorplan}). The results in Figure~\ref{fig:breathing_rates} show that, since \sys{} leverages the standard broadcast nature of WiFi networks, different users do not interfere with each other. Higher breathing rates are slightly more challenging to capture due to their corresponding required higher sampling rate. However, within the normal human breathing rate, \sys{} can achieve a high accuracy of 0.9 bpm error.

\subsection{Apnea Detection}
In this section, we investigate the system's performance in detecting apnea, i.e. absence of the breathing signal. The distance between the device and the AP is set to $2 \textrm{m}$ in this experiment. The users are requested to simulate the breathing pattern in Figure~\ref{fig:apneustic}, following the ground truth pattern with the help of our custom-made breathing metronome application.

\subsubsection{Impact of detection threshold ratio ($\tau$)}
\begin{figure}
  \centering
  \begin{subfigure}[t]{0.37\textwidth} 
                \centering
                \includegraphics[width=\textwidth]{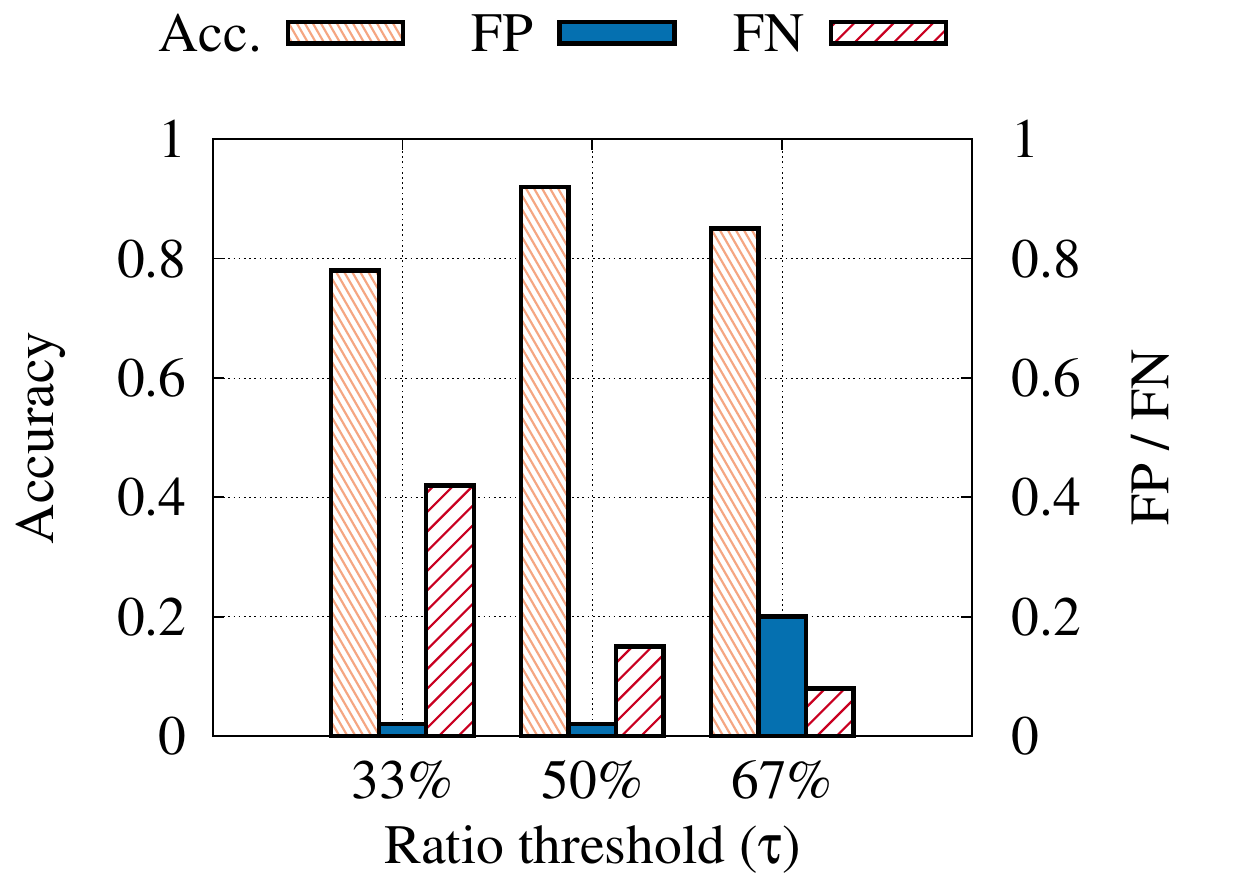}\\
                \caption{Holding device.}
                \label{fig:apnea_hold_device}
        \end{subfigure}
	\begin{subfigure}[t]{0.37\textwidth} 
                \centering
                \includegraphics[width=\textwidth]{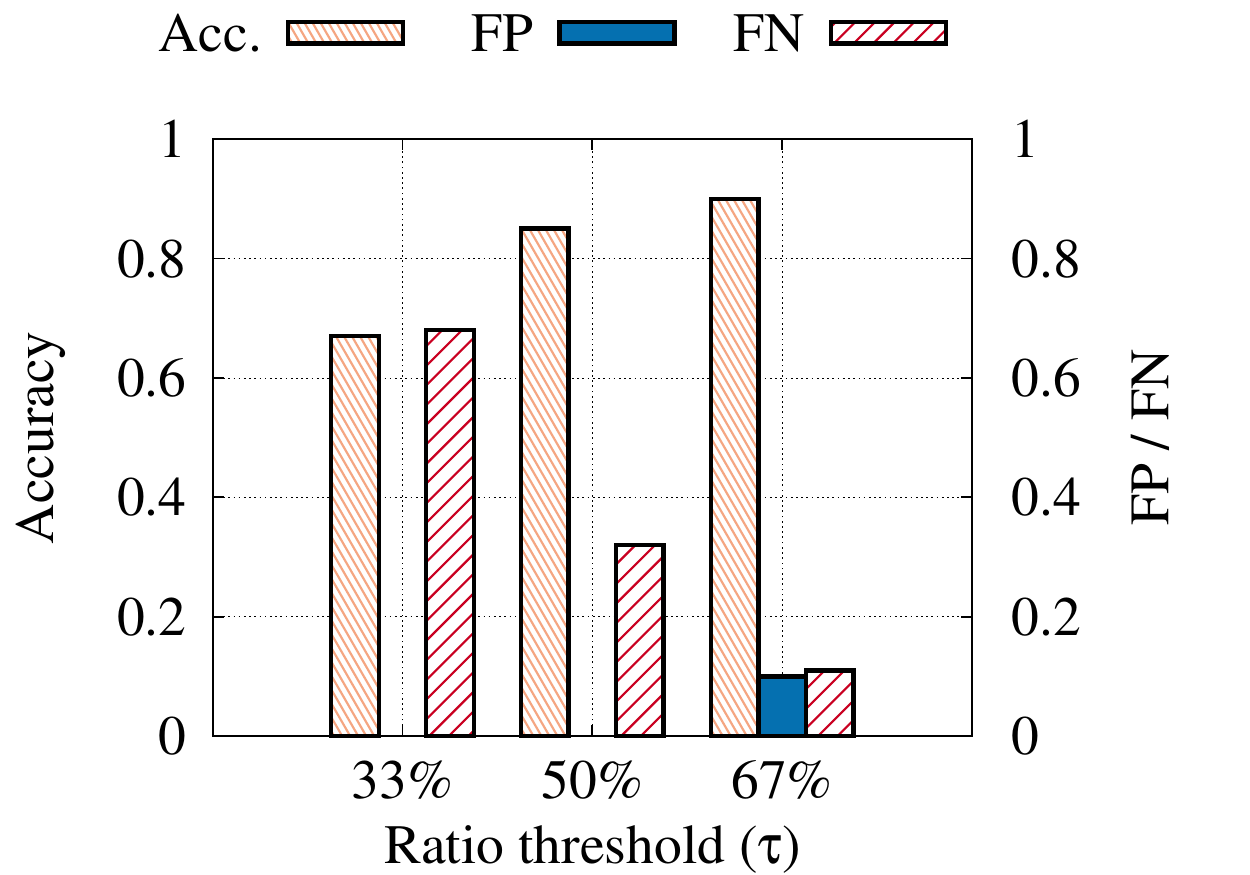}
                \caption{Hands-free.}
                \label{fig:apnea_los}
   \end{subfigure}

  \caption{Apnea detection accuracy for the two modes of operation using a single stream.}\label{fig:apea_res}
  \label{fig:apnea_detection}
   \vspace{-0.1in}
\end{figure}

Figure~\ref{fig:apnea_detection} shows the effect of the detection threshold ratio ($\tau$) on accuracy ($\textrm{True positives+True negatives} / {\textrm{Total samples}}$), false positive rate, and false negative rate. We observe that, as expected, increasing the detection threshold ratio ($\tau$) reduces the false negative rate while increasing the false positive rate. A value of $\tau$ between $50\%$ to $67\%$ leads to the best accuracy of 92\% for both the device-on-chest and hands-free scenarios. 

\subsubsection{Impact of multiple APs}
\begin{figure}
  \centering
  \begin{subfigure}[t]{0.37\textwidth} 
                \centering
                \includegraphics[width=\textwidth]{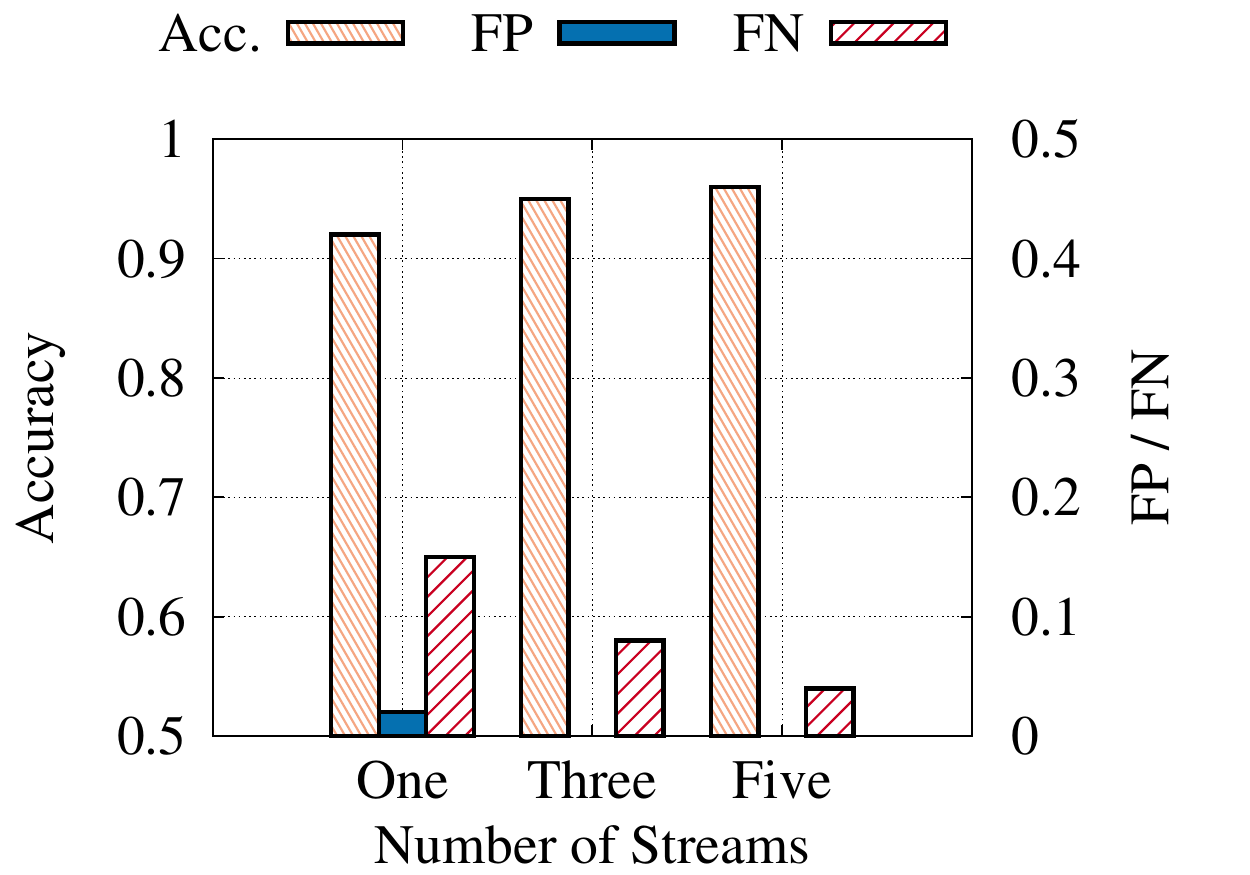}\\
                \caption{Holding device.}
                \label{fig:apnea_hold_device}
        \end{subfigure}
	\begin{subfigure}[t]{0.37\textwidth} 
                \centering
                \includegraphics[width=\textwidth]{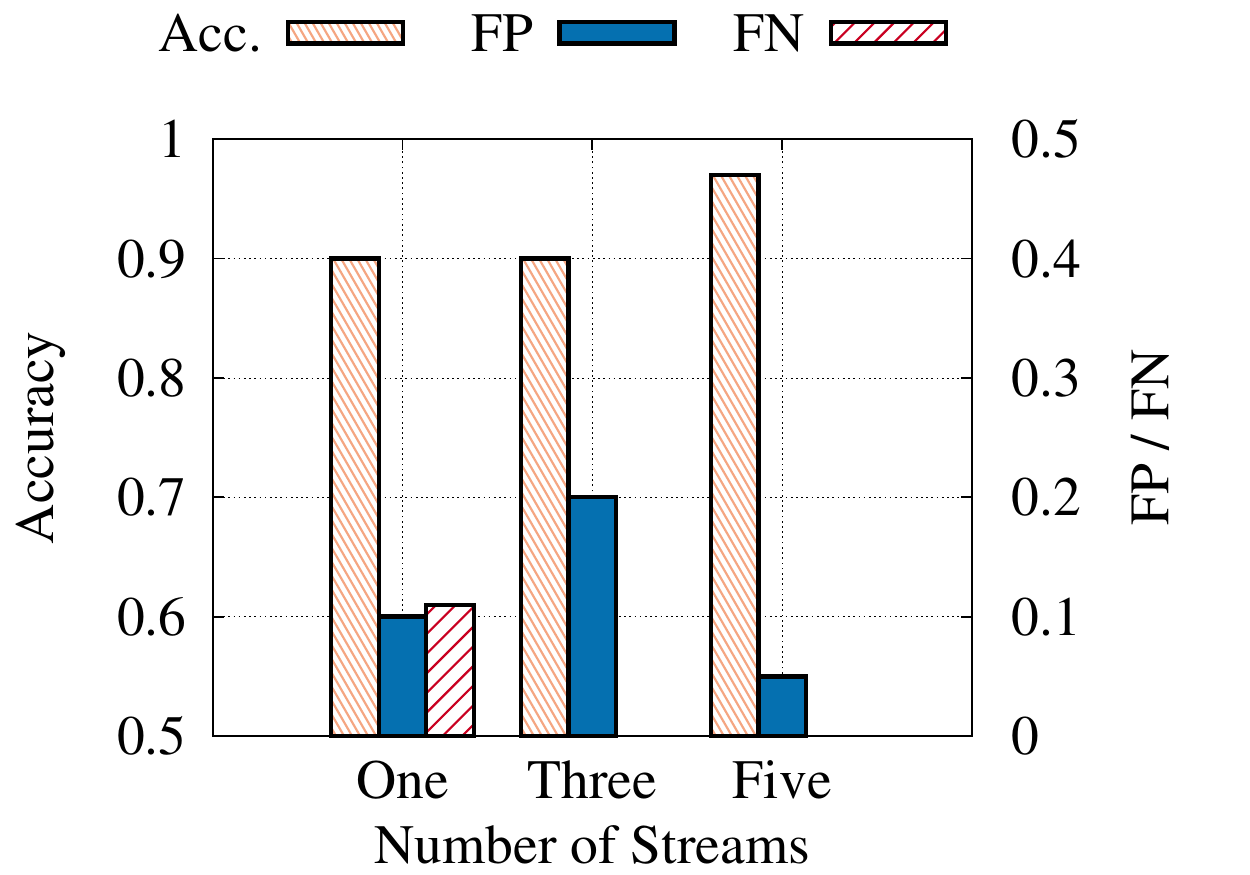}
                \caption{Hands-free.}
                \label{fig:apnea_los}
   \end{subfigure}

  \caption{Apnea detection accuracy for the two modes of operation using multiple streams.}\label{fig:apea_res}
  \label{fig:apnea_detection_multistream}
  \vspace{-0.1in}
\end{figure}

Figure~\ref{fig:apnea_detection_multistream} shows the effect of using more than one stream on apnea detection accuracy in the two modes of operation. A majority vote between all streams is used to select the correct state (apnea/normal breathing). The figure shows that increasing the number of streams generally increases the apnea detection performance under all metrics. Using five streams, \sys{} can achieve at least $96\%$ accuracy and 5\% false-negative and false positive rates. 

\subsection{Summary}
The results in this section show that \sys{}, through its different modules, can achieve a high accuracy of breathing rate extraction with an error less than 1 bpm in different scenarios with interfering humans performing their normal daily activities. This performance accuracy is robust to distances between the AP and device up to 8 m in through-the-wall scenario and 11 m for the no-wall scenario. The front user orientation provides the best accuracy due to the higher SNR. Other orientations still give a decent accuracy, with less than 2.6 bpm error

\sys{} can achieve a robust single time reading from its \emph{Robustness Enhancer} sub-module within 50 sec in all scenarios using the default parameters. This initial latency is useful for some applications, e.g. logging in the patient's chart, and does not affect the latency of the other system modules that \emph{all run in realtime}.

The false-negative rate is critical for apnea detection. Our results show that using the WiFi RSS streams from multiple overheard APs decrease the apnea detection false-negative rate to only 4\% in the on-chest scenario and zero in the hands-free scenario.

\section{Conclusion and Future Work}
\label{conclusion}

We presented \sys{}: a system for ubiquitous non-intrusive breathing rate monitoring based on the standard WiFi equipment. \sys{} combines a number of modules to extract the breathing signal from the noisy wireless channel as well as handle abrupt user actions and interfering humans. Implementation of \sys{} on different off-the-shelf WiFi devices showed that it can achieve a high accuracy of less than 1 bpm error under different realistic deployments in the device-on-chest and hands-free scenarios. Moreover, it can provide a robust reading for the patient's records in less than 50 seconds. \sys{}, in addition, can detect apnea with more than 96\% accuracy with less than 5\% false positive and negative rates. This allows it to be used in diverse scenario including remote patient monitoring, automatic early health-issues detection, fitness monitoring, infants sleeping monitoring, among others.

Currently, we are expanding \sys{} in different directions including detecting the exact apnea types, and performing clinical trials with real patients.

\bibliographystyle{abbrv}

\end{document}